\documentclass{emulateapj}
\usepackage{graphicx}% Include figure files
\usepackage{dcolumn}% Align table columns on decimal point
\usepackage{bm}% bold math
\usepackage{amsfonts}% bold math
\usepackage{color}
\usepackage{amsmath}
\usepackage{appendix}

\begin{document}

\title{
Magnetorotational Explosion of A Massive Star Supported by Neutrino Heating in General Relativistic Three Dimensional Simulations
}

\author{
Takami Kuroda$^{1}$,
Almudena Arcones$^{1,2}$,
Tomoya Takiwaki$^{3}$, and
Kei Kotake$^{4}$
}
\shorttitle{3D-GRMHD CCSN simulations with M1 neutrino transport}
\shortauthors{Kuroda et al.}

\affil{$^1$Institut f{\"u}r Kernphysik, Technische Universit{\"a}t Darmstadt, Schlossgartenstrasse 2, D-64289 Darmstadt, Germany\\
$^2$GSI Helmholtzzentrum f\"ur Schwerionenforschung, Planckstrasse 1, D-64291 Darmstadt, Germany\\
$^3$Division of Science, National Astronomical Observatory of Japan, 2-21-1, Osawa, Mitaka, Tokyo, 181-8588, Japan\\
$^4$Department of Applied Physics \& Research Institute of Stellar Explosive Phenomena, Fukuoka University, 8-19-1, Jonan, Nanakuma, Fukuoka, 814-0180, Japan}

\begin{abstract}
We present results of three-dimensional (3D), radiation-magnetohydrodynamics (MHD) simulations of core-collapse supernovae in full general relativity (GR) with spectral neutrino transport. 
In order to study the effects of progenitor's rotation and magnetic fields, we compute three models, where the precollapse rotation rate and magnetic fields are included parametrically to a 20 M$_{\odot}$ star.
While we find no shock revival in our two non-magnetized models during our simulation times ($\sim500$ ms after bounce), the magnetorotationally (MR) driven shock expansion immediately initiates after bounce in our rapidly rotating and strongly magnetized model. We show that the expansion of the MR-driven flows toward the polar directions is predominantly driven by the magnetic pressure, whereas the shock expansion toward the equatorial direction is supported by neutrino heating.
Our detailed analysis indicates that the growth of the so-called kink instability may hinder the collimation of jets, resulting in the formation of broader outflows.
Furthermore we find a dipole emission of lepton number, only in the MR explosion model, whose asymmetry is consistent with the explosion morphology. Although it is similar to the lepton-number emission self-sustained asymmetry (LESA), our analysis shows that the dipole emission occurs not from the protoneutron star convection zone but from above the neutrino sphere indicating that it is not associated with the LESA. We also report several unique neutrino signatures, which are significantly dependent on both the time and the viewing angle, if observed, possibly providing a rich information regarding the onset of the MR-driven explosion.
\end{abstract}


\section{Introduction}

\label{sec:Introduction}
The best-studied mechanism to explode massive stars ($\gtrsim 8 M_{\odot}$) is the neutrino mechanism \citep{Wilson85,Bethe85}, where neutrinos emitted from the protoneutron star (PNS) heat the matter behind the stalled bounce shock, leading to the shock revival into explosion, i.e., the onset of core-collapse supernova (CCSN) \citep[see, ][for reviews]{Radice18,Janka16,BMullerReview16,Foglizzo15,burrows13,Kotake12}. However, the neutrino mechanism generally fails in spherically symmetric (1D) simulations \citep[e.g.][]{Liebendorfer01,Sumiyoshi05} except for super-AGB stars \citep{kitaura} that cover the low-mass end of CCSN progenitors.

Multi-dimensional (multi-D) hydrodynamics has dramatic impacts on the neutrino mechanism \citep[see, e.g.,][]{Melson15b,Lentz15,Takiwaki16,BMuller17,Pan18,Ott18,O'Connor18,Summa18,Burrows19,Vartanyan19}.
Multi-D instabilities such as neutrino-driven convection and the standing accretion shock instability (SASI) \citep{Scheck06,Foglizzo06}, increase the dwell time of matter in the post-shock region, which substantially enhances the neutrino heating efficiency behind the shock.
Turbulence also plays a key role, providing the pressure support and energy transport 
 in the postshock region \citep[e.g.,][]{Abdikamalov15,Couch15,BMuller15,Roberts16,Takiwaki16,Radice18,Burrows19,Nagakura19}.
Other possible candidates to foster the onset of neutrino-driven explosions include inhomogenities in the progenitor's burning shells \citep[e.g.,][]{Couch15,BMuller17,yoshida19}, PNS convection \citep[see, e.g.,][]{Powell19,Nagakura20}, updates in neutrino opacities 
\citep[e.g., ][]{bollig17,Kotake18}, sophistication of neutrino transport schemes \citep[e.g., ][]{Sumiyoshi12,just18,Nagakura19}, and rotation and magnetic fields.
We focus on the final facet in this paper.

A number of effects of rotation in full 3D were first studied by \cite{Fryer04}, in which they explored the rotational effects on, e.g., the rotational instabilities, magnetic field amplification, and explosion dynamics.
Positive effects of rotation in favor of the onset of explosion include the larger shock radius due to the centrifugal force \citep{Nakamura3D14}, vigorous spiral SASI activity \citep{Summa18}, and 
energy transport via the rotational instability \citep{Takiwaki16}. On the 
other hand, rotation weakens the explodability because it results in a more extended and cooler PNS, which reduces the neutrino luminosities and energies \citep{Marek09}. These studies show that the impact of rotation on the neutrino mechanism depends sensitively on the precollapse rotation rate.
Supported by the outcomes from these state-of-the-art multi-D simulations, we are now reaching a broad consensus that the multi-D neutrino mechanism is the most promising way to account for canonical CCSNe with the explosion energies of the order of $10^{51}$ erg ($\equiv$ 1 Bethe, 1 B in short) or less.

The neutrino mechanism, however, is likely to fail in a subclass of CCSNe with very energetic
 explosion of $\sim$ 10 B, which is termed as hypernova (HN) \citep{iwamoto98}.
Observationally a HN is associated with collapse of very massive star typically with $\gtrsim30-40\ M_\odot$ in the main sequence stage \citep{Tanaka09}.
The magnetorotationally-driven mechanism originally proposed in the 1970s \citep{Bisnovatyi-Kogan70,LeBlancWilson70,Meier76,EMuller79} has received considerable attention.
The magnetorotational (MR) explosion mechanism relies on the extraction of the rotational free energy from the
central compact objects via the magnetic fields \citep[see also][in various contexts]{Blandford77,McKinney06}.

Rapid rotation of the iron core is a necessary condition for the working of the MR mechanism \citep[see][for collective references of early studies therein]{Kotake06}. In the collapsing core, the magnetic fields are amplified to dynamically relevant field strengths by rotational winding and/or magnetorotational instability (MRI) \citep{Akiyama03,Obergaulinger09,Masada15,Rembiasz16a}.
After bounce, the strong magnetic pressure launches the jets along the rotational axis \citep{arde00,Burrows07,Takiwaki09,Scheidegger10,Winteler12,Moesta14,Obergaulinger14}.
The highly aspherical explosion is also observationally supported by the analysis of the line
 profiles \citep[e.g.,][]{Maeda08}. Note in the non-rotating progenitors, \cite{Obergaulinger14} 
 were the first to point out that magnetorotationally-driven pressure support in the gain region (via turbulence) fosters the onset of neutrino-driven explosion. This result clearly presents evidence that implementation of 
sophisticated neutrino transport is needed for a quantitative study of magnetorotationally-driven CCSN modeling.

In the context of purely neutrino-driven models (without magnetic fields), it becomes certain that two-dimensional (2D) simulations overestimate the explodability for a wide variety of progenitor (\cite{Hanke12,Hanke13,Couch13,Dolence13,Takiwaki14}).  In order to correctly capture the evolution  and dynamics of the postshock turbulence, three-dimensional (3D) modeling is required. The higher explodability in 2D is also reported in MR models. \cite{Moesta14} has shown that a full 3D model leads to the formation of the less collimated (bipolar) jets than those in the counterpart octant symmetry model, which mimics 2D.
They pointed out that the less collimated outflow in 3D is an outcome of the so-called $|m|=1$ kink instability \citep{Lyubarskii99,Begelman98,Narayan09}. It has been demonstrated that the kink instability displaces the jet center from the rotational axis and prevents the magnetic fields amplification preferentially on the axis \citep[see also][]{Li00}.
More recently, \cite{Obergaulinger19} has reported the first 3D special-relativistic MHD simulations with spectral neutrino transport.
Their 3D models showed slightly longer explosion times, although the explosion energy and ejecta mass were higher and larger, respectively, compared to those in the counterpart 2D models.
Any remarkable non-axisymmetic instabilities, including the kink instability, were not seen in the 3D models of \citet{Obergaulinger19},  which is in contrast with \citet{Moesta14}. Therefore the multi-D effects in MHD models are still controversial, due partly to the limited number of full 3D MHD CCSN simulations reported so far \citep{Mikami08,Scheidegger10,Moesta14,Obergaulinger19}. 

% Kotake "I think that the following paragraph should be included in the next paper with the Almudena's group!"
%
%Multi-D effects also influence on the nucleosynthesis in the MHD CCSNe.
%The MHD explosion has been considered one of promising sites of $r$-process nucleosynthesis \citep{Winteler12,Nishimura15,Moesta18}, which is also supported by galactic chemical evolution simulations \citep[e.g.,][]{Cote19}.
%Regarding the $r$-process abundance pattern, however, there seems to be a strong dependence on 2D and 3D difference.
%\cite{Moesta18}, albeit with simple neutrino transport, showed that models with almost no influence of non-axisymmetric instabilities (i.e., essentially 2D models) can reach the third $r$-process peak, while the others with substantial influences can only up to the second peak.
%This is because that the less collimated jet and thus longer explosion time in non-axisymmetric full 3D model increase the dwell time of ejecta in the vicinity of PNS.
%It results in a relatively proton-rich ejecta due to absorption of more electron type neutrinos than 2D models.
%Therefore, the full 3D models showed basically no third peak $r$-process elements \citep{Moesta18}, while those elements can be produced in the counterpart 2D models \citep{Winteler12,Nishimura15,Moesta18}.
%For more detailed nucleosynthesis calculation, self-consistent full 3D GRMHD simulations with spectral neutrino transport are surely necessary (we will discuss the nucleosynthesis in our forthcoming paper). 

In this paper, we report first results of full 3D-GR, magnetorotational core-collapse simulations of a 20 $M_\odot$ star with spectral neutrino transport. 
We calculate three models, rotating magnetized, rotating non-magnetized, and non-rotating non-magnetized models.
Our results show that the MR explosion occurs in the rotating magnetized model shortly after core bounce, whereas the shock revival is not obtained in both non-magnetized models during our simulation time ($\sim500$ ms after bounce). While our results basically confirm the previous results \citep{Moesta14}, our  findings include detailed analysis of the kink instability, the dipole emission of lepton number in the MR explosion, and the neutrino signals from the 3D-GR MHD models with self-consistent neutrino transport.

This paper is organized as follows. Section \ref{sec:Numerical Methods and Computational Setup} starts with a concise summary of our GR-MHD neutrino transport scheme, which is followed by the initial setup of the simulation.
The main results and detailed comparison with previous studies are presented in Section \ref{sec:Results}.
We summarize our results and conclusions in Section \ref{sec:Discussion and Conclusions}.
Note that the geometrized unit is used in Section \ref{sec:Numerical Methods and Computational Setup},
i.e., the speed of light, the gravitational constant and the Planck constant are set to unity: $c= G = h=1$,
and cgs unit is used in Section \ref{sec:Results}.
The metric signature is $(-,+,+,+)$.
Greek indices run from 0 to 3 and Latin indices from 1 to 3, except $\nu$ and $\varepsilon$ denoting neutrino species and energy, respectively.
We also use a conventional expression for spatial coordinates $(x^1, x^2,x^3)=(x,y,z)$.

\section{Numerical Methods and Computational Setup}
\label{sec:Numerical Methods and Computational Setup}
In our full GR radiation-MHD simulations, we solve the evolution equations of metric, magnetohydrodynamics, and energy-dependent neutrino radiation.
Each of the evolution equations is solved in an operator-splitting manner, while the system evolves self-consistently as a whole satisfying the Hamiltonian and momentum constraints \citep{KurodaT12,KurodaT14,KurodaT16}.

\subsection{Basic $\nu$-GRMHD Equations}
\label{sec:Basic nu-GRMHD Equations}
Regarding the metric evolution, we evolve the standard BSSN variables $\tilde\gamma_{ij}$, $w(=e^{-2\phi})$ \citep{Marronetti08}, $\tilde A_{ij}$, $K$, and
$\tilde\Gamma^{i}$ \citep{Shibata95,Baumgarte99}.
Here $\phi\equiv \rm{ln}(\gamma)/12$ with $\gamma={\rm det}(\gamma_{ij})$.
The gauge is specified by the ``1+log'' lapse and by the Gamma-driver-shift condition.
Evolution equation of these variables are solved with a fourth-order finite difference scheme in space \citep{Zlochower05} and with a fourth-order Runge-Kutta time integration.
In appendix \ref{sec:Gowdy}, we show results of the polarized Gowdy wave test \citep{Alcubierre04} to show the fourth-order convergence of our metric solver.

In the radiation-magnetohydrodynamic part, the total stress-energy tensor $T^{\alpha\beta}_{\rm (total)}$ is expressed as
\begin{equation}
T_{\rm (total)}^{\alpha\beta} = T_{\rm (matter)}^{\alpha\beta} +T_{\rm (EM)}^{\alpha\beta}+\sum_{\nu\in\nu_e,\bar\nu_e,\nu_x}\int d\varepsilon T_{(\nu,\varepsilon)}^{\alpha\beta},
\label{TotalSETensor}
\end{equation}
where $T_{\rm (matter)}^{\alpha\beta}$, $T_{\rm (EM)}^{\alpha\beta}$, and $T_{(\nu,\varepsilon)}^{\alpha\beta}$ are the 
stress-energy tensor of matter, electro-magnetic, and energy $(\varepsilon)$ dependent neutrino radiation field of specie $\nu$, respectively.
We consider all three flavors of neutrinos ($\nu_e,\bar\nu_e,\nu_x$) with $\nu_x$ representing heavy-lepton neutrinos (i.e. $\nu_{\mu}, \nu_{\tau}$ and their anti-particles).
$\varepsilon$ represents the neutrino energy measured in the 
comoving frame.
In this paper, although we omit to describe detailed evolution equations of the neutrino radiation field \citep[we refer the reader to][]{KurodaT16}, we solve spectral neutrino transport of the zeroth and first order radiation momenta, based on the truncated moment formalism \citep{Shibata11} employing an M1 analytical closure scheme.

In the following, we briefly describe our GRMHD formulation.
The stress-energy tensor of electro-magnetic field $T_{\rm (EM)}^{\alpha\beta}$ is expressed as
\begin{eqnarray}
T_{\rm (EM)}^{\alpha\beta}=F^{\alpha\delta}F^{\beta}_{\delta}-\frac{1}{4}g^{\alpha\beta}F_{\delta\gamma}F^{\delta\gamma},
\end{eqnarray}
where $F^{\alpha\beta}$ is the electromagnetic field tensor.
Since we currently consider the ideal MHD case, Maxwell's equations are written in terms of the dual tensor $F^\ast_{\alpha\beta}=\frac{1}{2}\epsilon_{\alpha\beta\gamma\delta}F^{\gamma\delta}$ as
\begin{eqnarray}
\label{eq:Maxwell}
\nabla_\beta F^{\ast \beta}_{\alpha}=0.
\end{eqnarray}
We define the magnetic field four vector $b^\alpha$ as below
\begin{eqnarray}
b^\alpha=-\frac{1}{2}\epsilon^{\alpha\beta\gamma\delta}u_\beta F_{\gamma\delta},
\end{eqnarray}
with $\epsilon^{\alpha\beta\gamma\delta}$ and $u_\alpha$ being the Levi-Civita tensor and matter four velocity, respectively.
In addition, for later convenience, the magnetic field three vector $B^i$ should also be introduced as 
\begin{eqnarray}
B^i\equiv F^{\ast it}=-{\gamma^i}_jn_\mu F^{\ast j\mu}= Wb^i-\alpha b^t u^i,
\end{eqnarray}
where $W=-u^\mu n_\mu$ is the Lorentz factor (do not confuse with $w=e^{-2\phi}$ of geometrical variables) and $n_\mu=(-\alpha,0,0,0)$ is a unit vector normal to the spacelike hypersurface foliated into the spacetime.
Then, using the orthogonality condition $B^\alpha n_\alpha=0$, the time and spacial components of Eq. (\ref{eq:Maxwell}) can be rewritten as
\begin{eqnarray}
\label{eq:solenoidal}
\partial_i(\sqrt{\gamma}B^i)=0,
\end{eqnarray}
i.e., the solenoidal constraint of $B^i$, and 
\begin{eqnarray}
\label{eq:induction}
\partial_t(\sqrt{\gamma}B^i)+\partial_j\sqrt{\gamma} (v^jB^i-B^jv^i)=0,
\end{eqnarray}
respectively, where  $v^i\equiv u^i/u^t$.

Additionally to the evolution equation (\ref{eq:induction}) of the magnetic field, we solve the following ideal hydrodynamic equations \citep[see, e.g.][]{Shibata05b,Gammie03} including electron number conservation
%{\setlength\arraycolsep{2pt}
\begin{eqnarray}
\label{eq:GRmass}
&&\partial_t \rho_{\ast}+\partial_i(\rho_\ast v^i)=0,\\
\label{eq:GRmomentum}
&&\partial_t \sqrt{\gamma} S_i+\partial_j \sqrt{\gamma}( S_i v^j+\alpha P_{\rm tot}\delta_i^j-\alpha B^j(B_i+B^ku_ku_i)/W^2)\nonumber \\
&&=-\sqrt{\gamma}\biggl[ S_0\partial_i \alpha- S_k\partial_i \beta^k-2\alpha S_k^k\partial_i \phi \nonumber \\
&&+\alpha e^{-4\phi} ({S}_{jk}-P_{\rm tot} \gamma_{jk}) \partial_i 
\tilde{\gamma}^{jk}/2%&&\nonumber \\
+\alpha \int d\varepsilon \sum_\nu S_{(\nu,\varepsilon)}^\mu \gamma_{i\mu} \biggr],\\%\nonumber\\ \\
\label{eq:GRenergy}
&&\partial_t \sqrt{\gamma} \tau+\partial_i \sqrt{\gamma}(\tau v^i+P_{\rm tot}(v^i+\beta^i)-\alpha B^jB^ku_k/W)=\nonumber \\
&&\sqrt{\gamma}\biggl[ \alpha K S_k^k /3+\alpha e^{-4\phi} ({S}_{ij}-P_{\rm tot} \gamma_{ij})\tilde{A^{ij}}  \nonumber\\
&&\ \ \ \ \ \ \ \ \ \ \ \ \ \ \ - S_iD^i\alpha+\alpha \int d\varepsilon \sum_\nu S_{(\nu,\varepsilon)}^\mu n_\mu \biggr],
\end{eqnarray}
and
\begin{eqnarray}
\label{eq:GRYe}
\partial_t (\rho_\ast Y_e)+\partial_i (\rho_\ast Y_e v^i)=\sqrt{\gamma}\alpha m_{\rm u}\int \frac{d\varepsilon}{\varepsilon}
(S_{(\nu_e,\varepsilon)}^\mu-S_{(\bar\nu_e,\varepsilon)}^\mu) u_\mu, \nonumber \\
\end{eqnarray}
%}
where $\rho_\ast=\rho \sqrt{\gamma}W$,
$S_i=(\rho h+b^2)W u_i-\alpha b^t b_i $, $S_{ij}=(\rho h+b^2) u_i u_j+P_{\rm tot}\gamma_{ij}-b_i b_j$, $S_k^k=\gamma^{ij}S_{ij}$, $\tau= S_0-\rho W$,
$S_0=(\rho h+b^2) W^2-P_{\rm tot}-(\alpha b^t)^2$.
On the right hand side of Eq.(\ref{eq:GRenergy}), $D^i$ represents the covariant derivative with respect to the three metric $\gamma_{ij}$.
$\rho$ is the rest mass density and $h= 1+e_{\rm mat}+P_{\rm mat}/\rho$ is the specific enthalpy of matter (composed of baryons, electrons, and photons) with $e_{\rm mat}$ and $P_{\rm mat}$ being the specific internal energy and pressure of matter, respectively.
$b^2=b^\alpha b_\alpha$, $P_{\rm tot}=P_{\rm mat}+P_{\rm mag}$ is the total pressure, $P_{\rm mag}=b^2/2$ is the magnetic pressure, $Y_e\equiv n_e/n_b$ is the electron fraction ($n_e$ and $n_b$ are the number densities of electrons and baryons, respectively), and $m_{\rm u}$ is the atomic mass unit.
$P_{\rm mat}(\rho,s,Y_e)$ and $e_{\rm mat}(\rho,s,Y_e)$ are given by an equation of state (EOS) 
with $s$ denoting the entropy per baryon.

We thus evolve the following magnetohydrodynamic and radiation conservative variables 
\begin{eqnarray}
\label{eq:conservativeV}
\bf Q=\left[
\begin{array}{c}
\rho^\ast \\
\sqrt{\gamma}S_i \\
\sqrt{\gamma}\tau  \\
\sqrt{\gamma} B^i \\
\rho^\ast Y_e \\
\sqrt{\gamma}E_{(\nu,\varepsilon)} \\
\sqrt{\gamma}{F_{(\nu,\varepsilon)}}_i \\
\end{array}
\right],
\end{eqnarray}
where $(E_{(\nu,\varepsilon)},{F_{(\nu,\varepsilon)}}_i)$ are the zeroth and first order moments of neutrino radiation \citep{Shibata11,KurodaT16}.

Every time we update the conservative variables $\bf Q$, we obtain the following primitive variables
\begin{eqnarray}
\label{eq:primitiveV}
\bf P=\left[
\begin{array}{c}
\rho \\
u^i \\
s \\
B^i \\
Y_e \\
E_{(\nu,\varepsilon)} \\
{F_{(\nu,\varepsilon)}}_i \\
\end{array}
\right]
\end{eqnarray}
by Newton's method.

\subsection{Constrained Transport}
\label{sec:Constrained Transport}
We solve the conservation equations (\ref{eq:GRmass})-(\ref{eq:GRYe}) using the HLL scheme \citep{Harten83}.
Meanwhile the induction equation (\ref{eq:induction}) is solved by a constrained transport (CT) method \citep{Evans&Hawley88} to satisfy the solenoidal condition Eq. (\ref{eq:solenoidal}).
For the CT method, we also utilize the HLL scheme when we reconstruct the electric field that will be mentioned shortly here.
To solve the (HLL) Riemann problem, we need to evaluate the left and right states at cell surface.
The left and right states are interpolated from cell centered primitive variables $\bf P$ and some of the metric terms $(w,\alpha,\beta^i,\gamma_{ij})$, which are needed to evaluate the full conservative variables $\bf Q$, by a spatial reconstruction.
We use the Piecewise Parabolic Method (PPM) for the spatial reconstruction (\cite{Colella84} or \cite{Hawke05} for more suitable upwind reconstruction method in GR.).
After the spatial reconstruction step, we calculate the fastest left- and right-going wave speeds \citep[e.g.][]{Anton06} and the HLL fluxes.

We also introduce the electric field $E^i$ defined by
\begin{eqnarray}
{\bf E}=\sqrt{\gamma}{\bf (v\times B)}
\end{eqnarray}
for the CT method.
Then the equation (\ref{eq:induction}) can be rewritten as
\begin{eqnarray}
\label{eq:rotE}
\partial_t(\sqrt{\gamma}B^i)-({\bf \nabla\times E})^i=0.
\end{eqnarray}
Employing a usual staggered mesh algorithm, we define the magnetic field $B^i$ and the electric field $E^i$ at cell surface and edge, respectively, while the rest of variables are defined at cell center.
For instance, $B^x$ and $E^x$ are defined at $(i+1/2,j,k)$ and $(i,j+1/2,k+1/2)$, respectively, where $(i,j,k)$ denotes the cell center and, e.g. $j+1/2$ represents a displaced position from cell center along $y$ axis with a half cell width.
As in the usual manner, the electric field $E^i$ defined on the cell edge is obtained from the HLL flux for $B^i$, corresponding to the advection term in Eq. (\ref{eq:induction}).
We use the nearest four electric fields defined on the cell surface, i.e., corresponding terms in the HLL flux, and take their simple arithmetic average\footnote{Although we used a simple arithmetic average in this study, we later found that the upwind reconstruction \citep[e.g.][]{Athena++} could significantly reduce numerical oscillations seen in the reconstructed electric field, especially outside the SN shock surface where the flow is supersonic, which eventually led to the crash of current MHD simulations.}.

%Here the cell centered magnetic field $(\equiv\hat B^i)$ is defined by a first order arithmetic average of the cell surfaced value $B^i$, e.g. $\hat B^x(i,j,k)\equiv \left(B^x(i+1/2,j,k)+B^x(i-1/2,j,k)\right)/2$.

%The original PPM method contains four reconstruction steps that are in order: interpolation, steepening, flattening, and preserving monotonicity.
%Among these four steps, the second and third steps contain contact and shock discontinuities finding that is judged by a certain condition of rest mass density and pressure, respectively.
%In contrast to the original PPM method which assumes a simple $\Gamma$-law EOS, our basic hydro equations contain the conservation of electrons and also use the entropy $s$ as a primitive variable for convenience.
%Here the left and right states of the magnetic field are of course identical to the corresponding $B^i$ defined on each cell surface and do not need to be reconstructed.

Our numerical grid employs a fixed nested structure and there is a boundary between different numerical resolutions.
Therefore we apply a refluxing procedure both for the HLL fluxes and the electric field $E^i$ \citep{KurodaT10} before solving Eqs.(\ref{eq:GRmass})-(\ref{eq:GRYe}) and (\ref{eq:rotE}) to satisfy the conservation law and solenoidal constraint in the whole computational domain.

\subsection{Initial Setup}
\label{sec:Initial Setup}
We study the frequently used solar-metallicity model of the 20 $M_{\odot}$ star ``s20a28n'' from \cite{WH07}.
Although one of our final aims is to understand the hypernova explosion mechanism of very massive stars $(\gtrsim30\ M_\odot)$, this progenitor star is widely used in previous studies \citep[e.g.,][]{Melson15b,Ott18,O'Connor18,Burrows19} and our non-rotating, non-magnetized model (see below) could thus be a reference model to calibrate our 3D $\nu$-GRMHD code.
For the nuclear EOS, we use SFHo of \cite{SFH}.
The 3D computational domain is a cubic box with $3\times10^4$ km width in which nested boxes with 10 refinement levels are embedded.
Each box contains $128^3$ cells and the minimum grid size near the origin is $\Delta x=458$ m.
In the vicinity of the stalled shock at a radius of $\sim100$ km, our resolution 
achieves $\Delta x\sim\ 1.9$ km, i.e., the effective angular resolution becomes $\sim1^\circ$.
The neutrino energy space $\varepsilon$ logarithmically covers from 1 to 300 MeV with 12 energy bins.
Regarding neutrino opacities, the standard weak interaction set in \citet{Bruenn85}, which are:
absorption and emission process
{\setlength\arraycolsep{2pt}
\begin{eqnarray}
\nu_en&\leftrightarrow& e^-p, \\
\bar\nu_ep&\leftrightarrow& e^+n,\\
\nu_eA&\leftrightarrow& e^-A',
\end{eqnarray}}
isoenergy scattering of neutrinos off nucleons and heavy nuclei
{\setlength\arraycolsep{2pt}
\begin{eqnarray}
\nu n&\leftrightarrow& \nu n,\\
\nu p&\leftrightarrow& \nu p,\\
\nu A&\leftrightarrow& \nu A,
\end{eqnarray}}
inelastic neutrino electron scattering
{\setlength\arraycolsep{2pt}
\begin{eqnarray}
\nu e&\leftrightarrow& \nu e,
\end{eqnarray}}
and thermal neutrino pair production and annihilation
{\setlength\arraycolsep{2pt}
\begin{eqnarray}
e^-e^+&\leftrightarrow& \nu \bar\nu
\end{eqnarray}}
are taken into account.
In addition, nucleon-nucleon Bremsstrahlung \citep{Hannestad98}
{\setlength\arraycolsep{2pt}
\begin{eqnarray}
NN&\leftrightarrow& NN\nu \bar\nu
\end{eqnarray}}
is also included \citep[for more details, see][]{KurodaT16}.

The original progenitor model ``s20a28n'' assumes neither rotation nor magnetic fields.
We thus artificially add them to the non-rotating progenitor model.
We employ a widely used cylindrical rotational profile \citep{Dimmelmeier02A}
\begin{equation}
    u^tu_\phi=\varpi_0^2(\Omega_0-\Omega),
\end{equation}
where $u_\phi\equiv\varpi^2\Omega$ with $\varpi=\sqrt{x^2+y^2}$.
$\Omega$ is the angular frequency of fluid element.
Using $\Omega$, the rotational component of the initial four velocity is simply set by $(u_x,u_y,u_z)=\Omega(y,x,0)$.
$\varpi_0$ and $\Omega_0$ indicate the size and angular frequency of a rigidly rotating central cylinder, respectively.
Note that $\Omega_0$ and $\Omega$ are measured by an Eulerian observer.
This rotational profile gives the angular frequency falling off with $\varpi^{-2}$ beyond $\varpi_0$, i.e., the specific angular momentum asymptotically reaches a constant value $\varpi_0^2\Omega_0$.

For the initial magnetic fields that should satisfy the solenoidal constraint, we use the following purely toroidal vector potential \begin{eqnarray}
A_\phi&=&\frac{B_0}{2}\frac{R_0^3}{r^3+R_0^3}r\sin{\theta},\\
A_r&=&A_\theta=0.
\end{eqnarray}
Here $(r,\theta,\phi)$ denote the usual coordinates in the spherical polar coordinate system.
By defining these vector potentials on the cell edge and taking their curl $\bf B=\nabla\times{\bf A}$, the magnetic field defined on the numerical cell surface automatically satisfies the solenoidal constraint.
This vector potential gives nearly uniform magnetic field parallel to the rotational axis (i.e. $z$-axis) for $r\lesssim R_0$ and dipolar magnetic field for $r\gtrsim R_0$.

We set $\varpi_0=R_0=10^8$ cm corresponding roughly to the iron core size at the precollapse stage.
We calculate three models: rotating magnetized, rotating non-magnetized, and non-rotating non-magnetized.
For the rotating models, we set $\Omega_0=1$ rad s$^{-1}$.
This value is very reasonable compared to the one of a rotating 20 $M_{\odot}$ model in \cite{Heger00} that gives $\Omega_0\sim3$ rad s$^{-1}$.
Regarding the magnetic field strength at origin, we set $B_0=10^{12}$ G that can be amplified strongly enough to affect the dynamics through simple linear amplification mechanisms, i.e., compression and rotational wrapping, during collapse and is also widely used in most of previous MHD simulations \citep{Burrows07,Takiwaki09,Scheidegger10,Moesta14,Obergaulinger19}. Three models are labeled as R0B00, R1B00, and R1B12, where the integer after R denotes $\Omega_0$.
B00 and B12 represent $B_0=0$ and $10^{12}$ G, respectively.

\section{Results}
\label{sec:Results}
In this Section, we explain our main results.
Sections \ref{sec:Postbounce Evolution} and \ref{sec:Shock Wave Evolution} are devoted to explaining general hydrodynamic properties in the post-bounce evolution.
In Section \ref{sec:Non-axisymmetric instabilities inside the MHD outflow}, we discuss non-axisymmetric instabilities in the PNS and MHD outflow, which is relevant to the neutrino signals in Section \ref{sec:Rotational Effects on Neutrino Profiles}. The role of neutrino heating in the MR mechanism is addressed in Section \ref{sec:The Role of Neutrino Heating}. We explain the dipole emission of lepton number in our MR explosion model in 
Section \ref{sec:The Asymmetry of Lepton Number Emission}.

\subsection{Postbounce Evolution}
\label{sec:Postbounce Evolution}
%%%%%%%%%%%%%%%%%%%%%%%%%%%%%%%%%%%%%%
% Memo for visit.
%R0B00    0<Tpb<100ms  280- 780, set Ent_max=12
%R0B00 100<Tpb<200ms 785-1280, set Ent_max=20
\begin{figure*}[bh]
\begin{center}
\includegraphics[width=50mm,angle=0.]{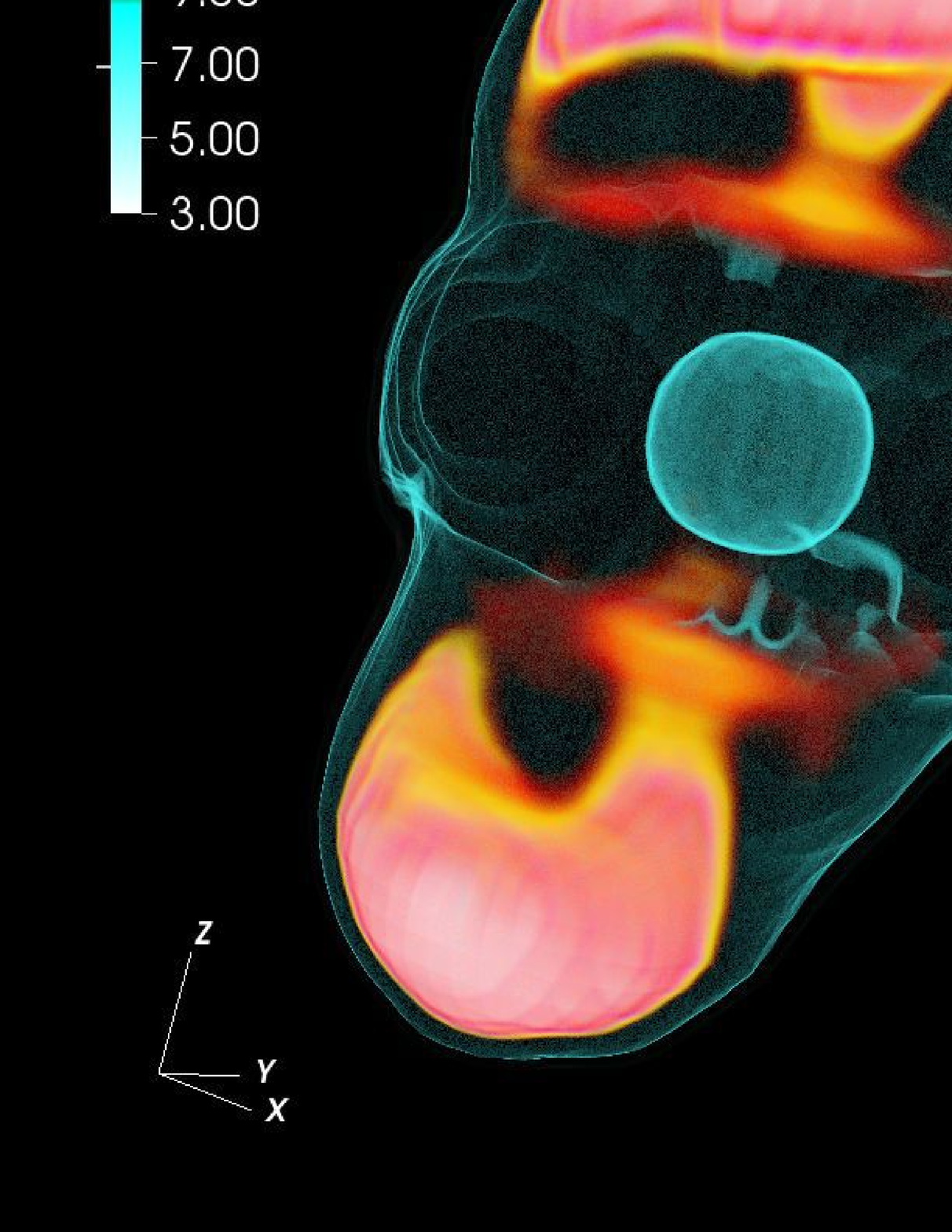}
\includegraphics[width=50mm,angle=0.]{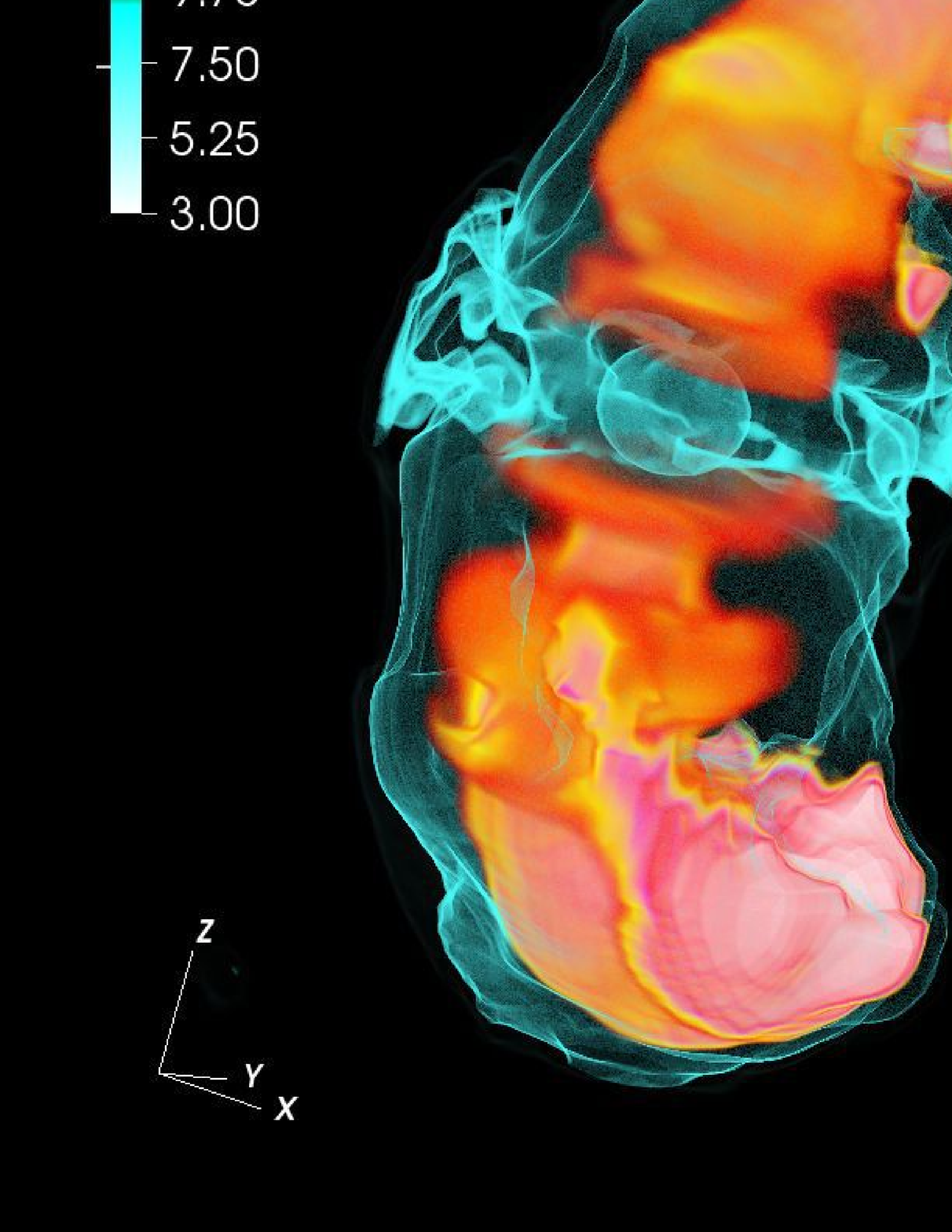}
\includegraphics[width=50mm,angle=0.]{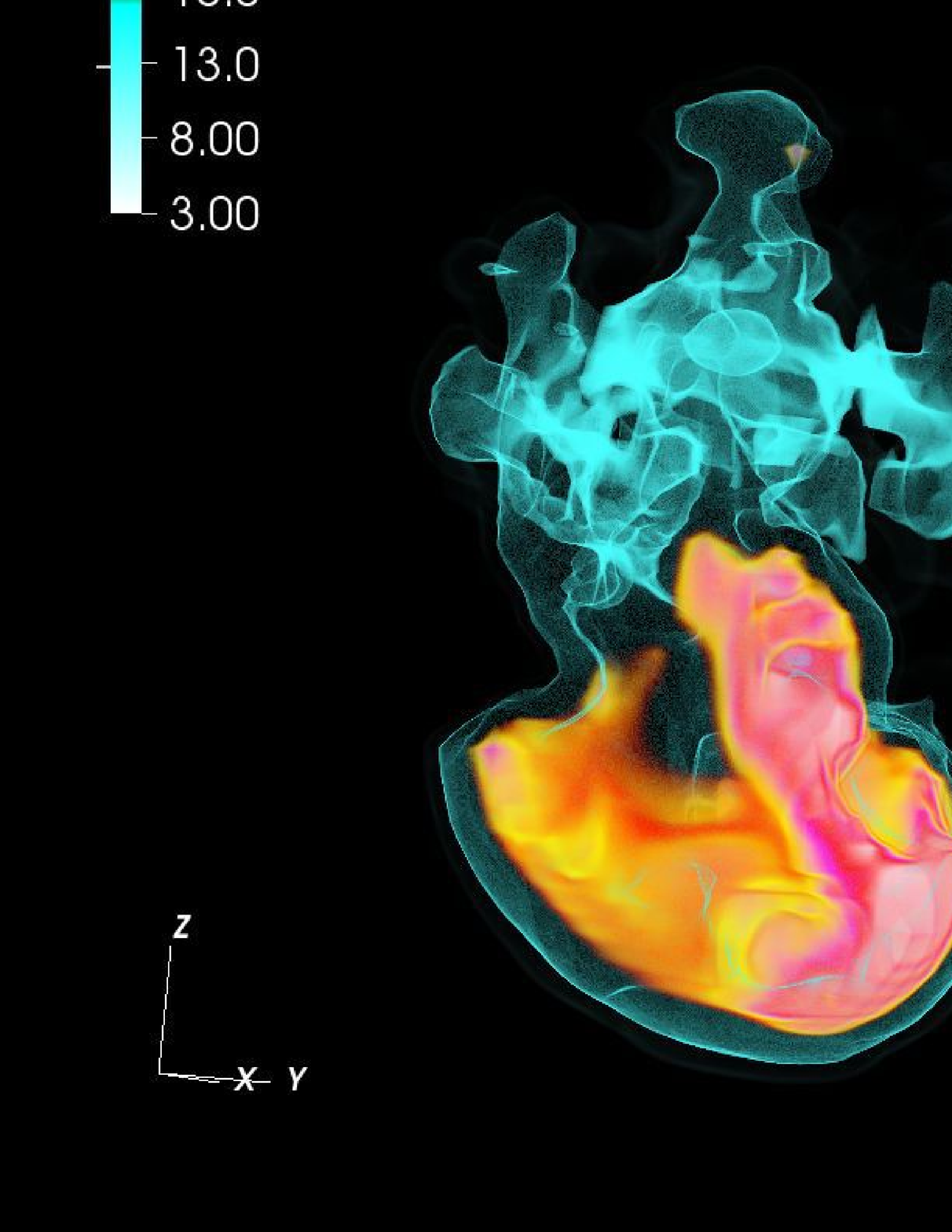}\\
\includegraphics[width=50mm,angle=0.]{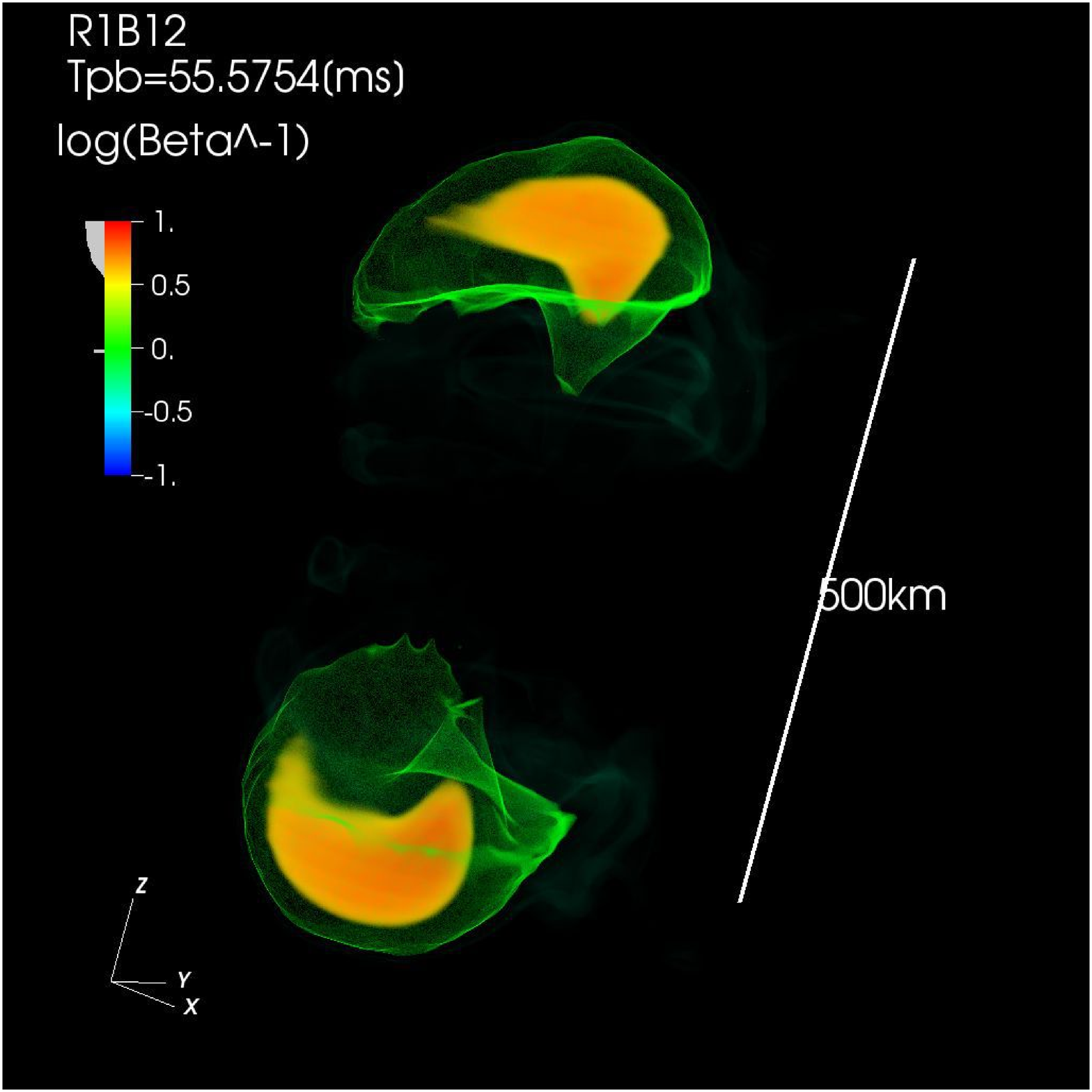}
\includegraphics[width=50mm,angle=0.]{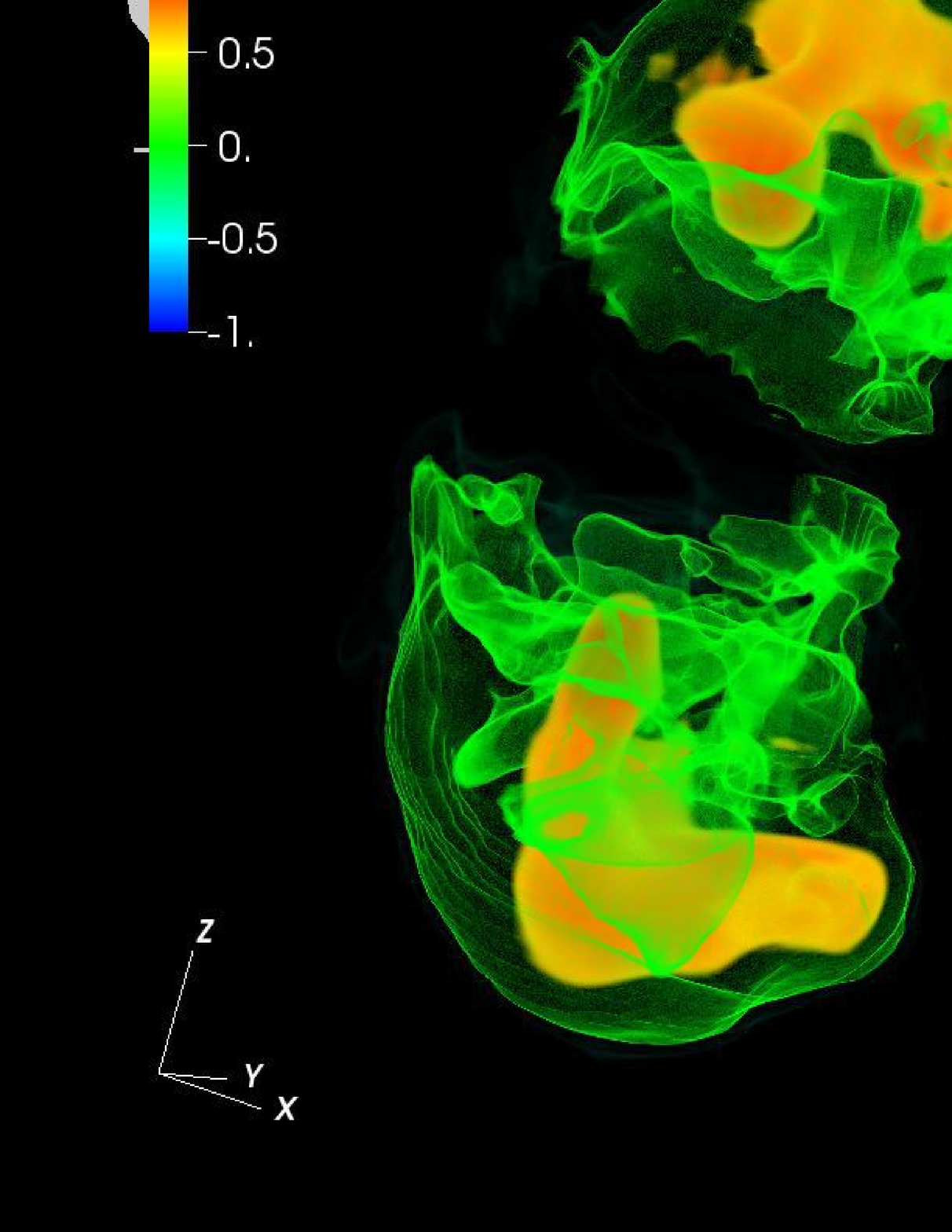}
\includegraphics[width=50mm,angle=0.]{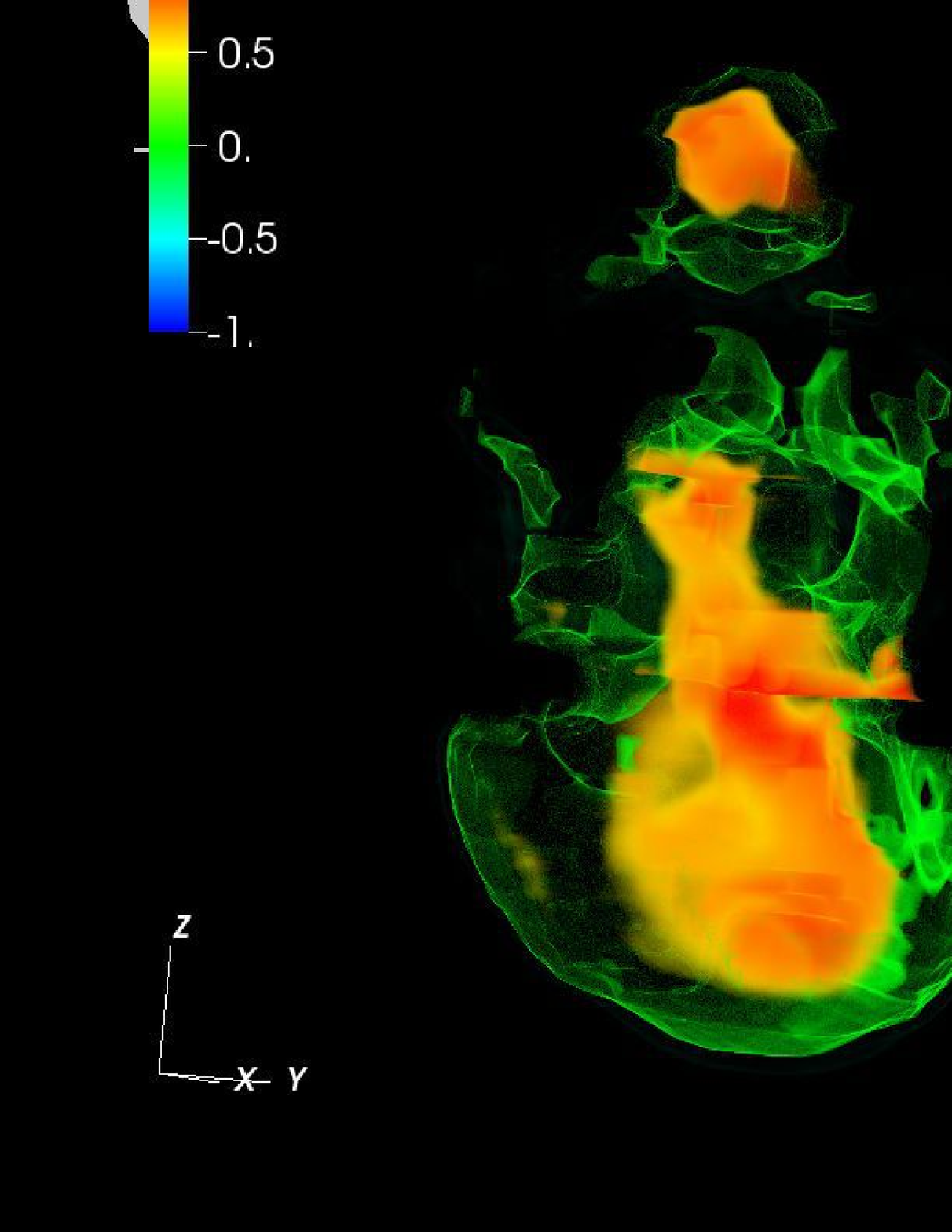}
  \caption{Snapshots of the volume rendered entropy (upper panels) and inverse of the plasma $\beta$ in the logarithmic scale ($\log\beta^{-1}$, lower panels) for model R1B12. From left to right panels, the post-bounce time of $t_{\rm pb}\sim56$ ms, 100 ms, and 250 ms are depicted, respectively. In the upper panels, the central bluish spherical/spheroidal object roughly corresponds to the unshocked PNS core. Note that the inclination angle of the coordinates is not fixed in each time snapshot to visualize the expansion morphology more clearly. The white line denotes the length scale that is parallel to the rotational axis ($z$-axis).
  \label{fig:R1B12_Ent_Beta}
}
\end{center}
\end{figure*}
%%%%%%%%%%%%%%%%%%%%%%%%%%%%%%%%%%%%%%
%%%%%%%%%%%%%%%%%%%%%%%%%%%%%%%%%%%%%%
\begin{figure*}[htbp]
\begin{center}
\includegraphics[width=50mm,angle=0.]{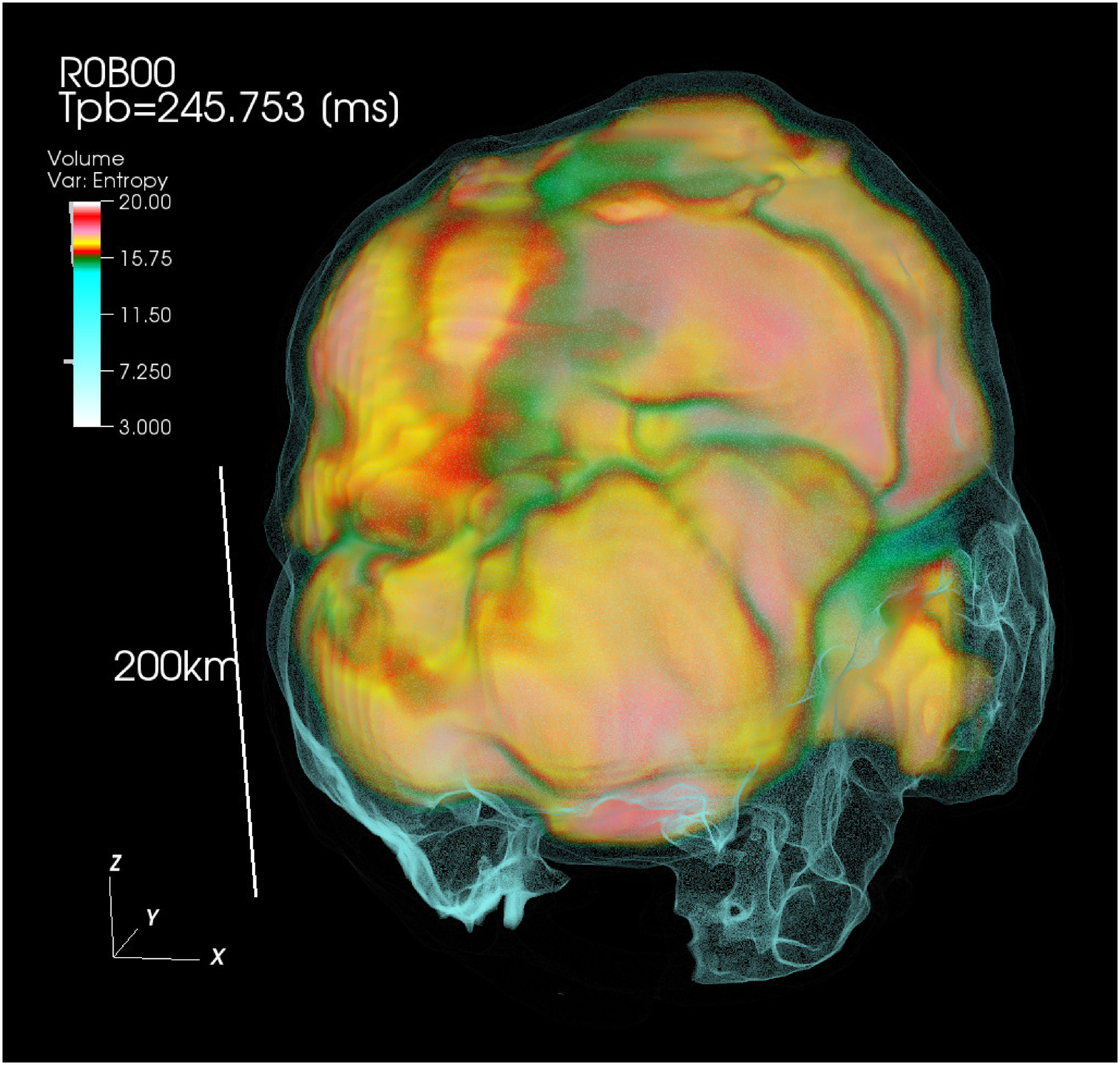}
\includegraphics[width=50mm,angle=0.]{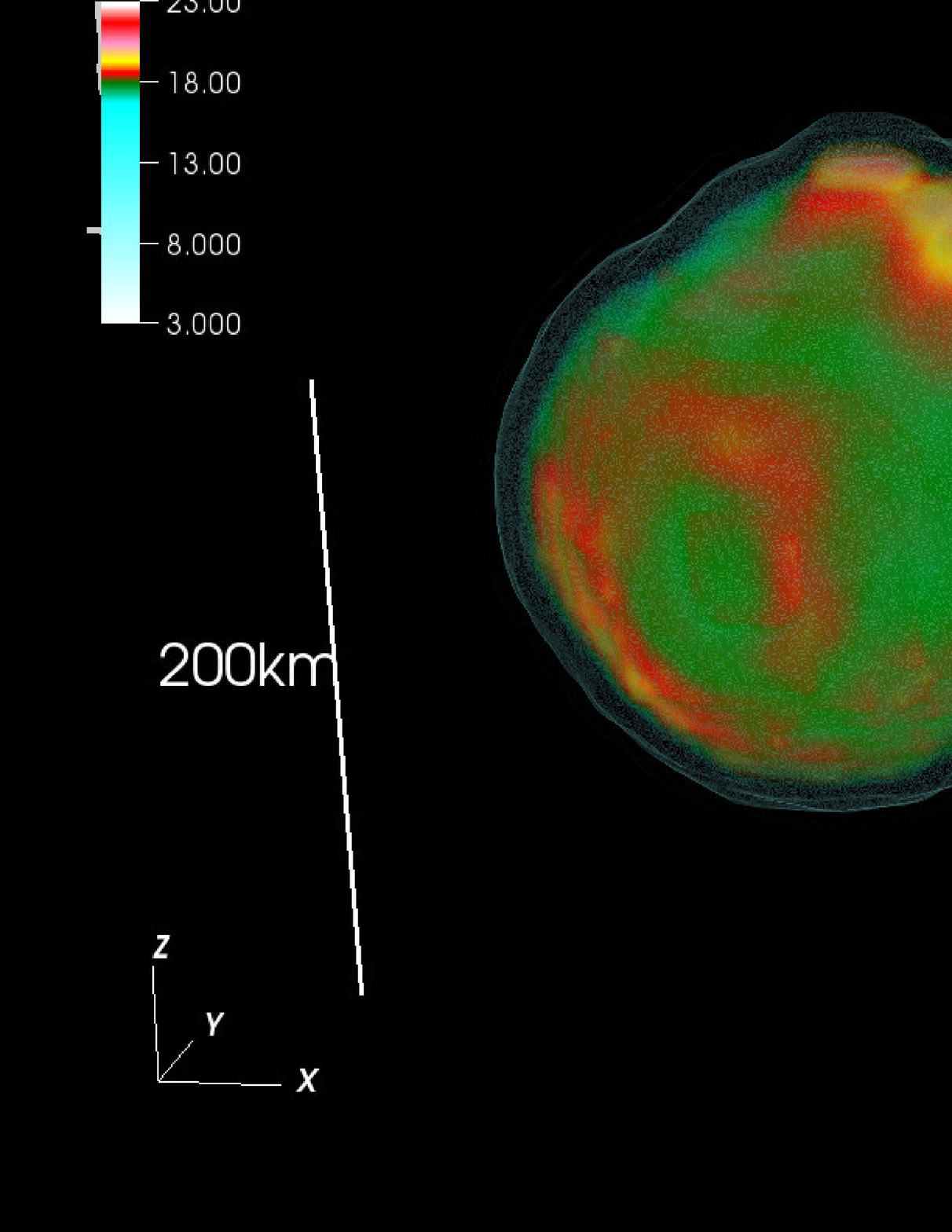}
\includegraphics[width=50mm,angle=0.]{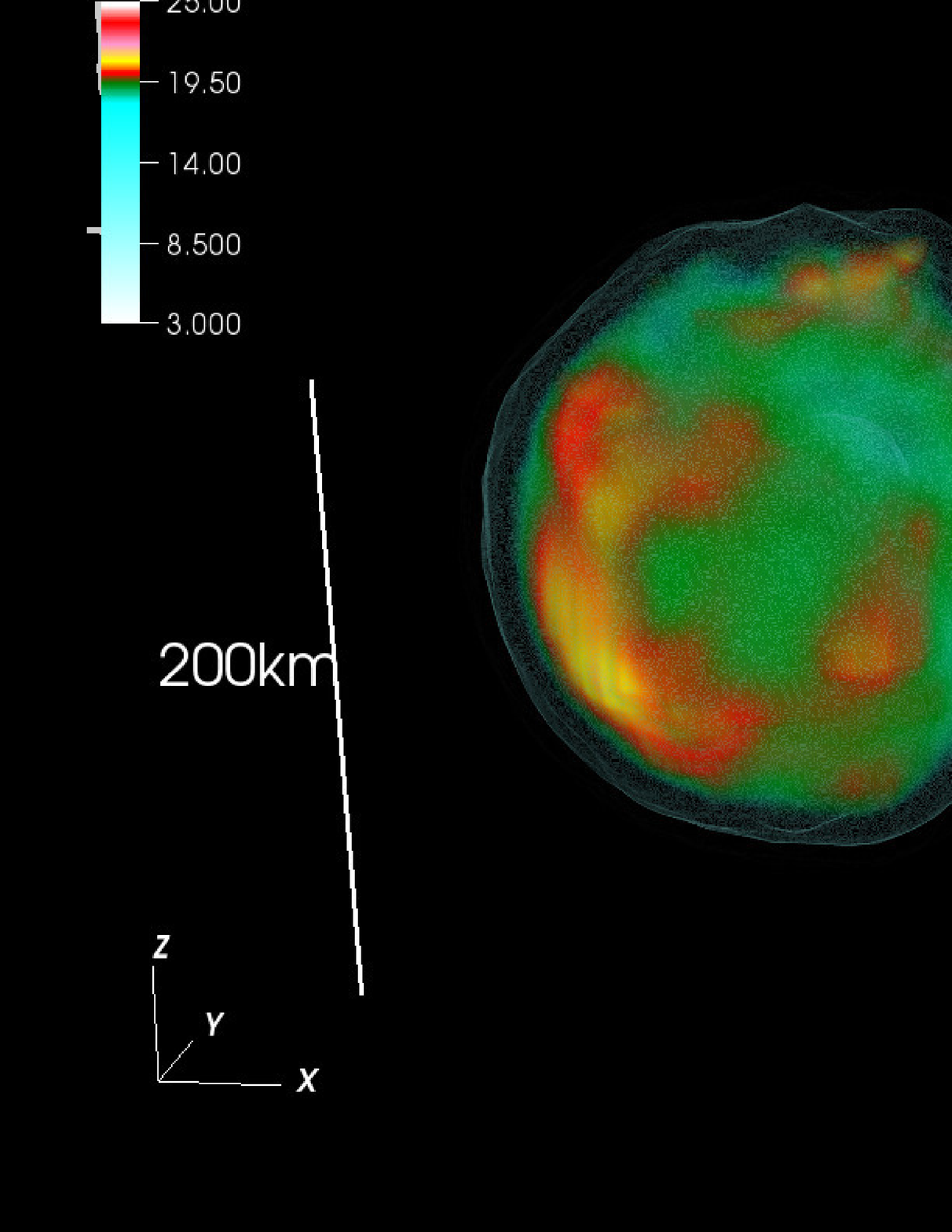}\\
\includegraphics[width=50mm,angle=0.]{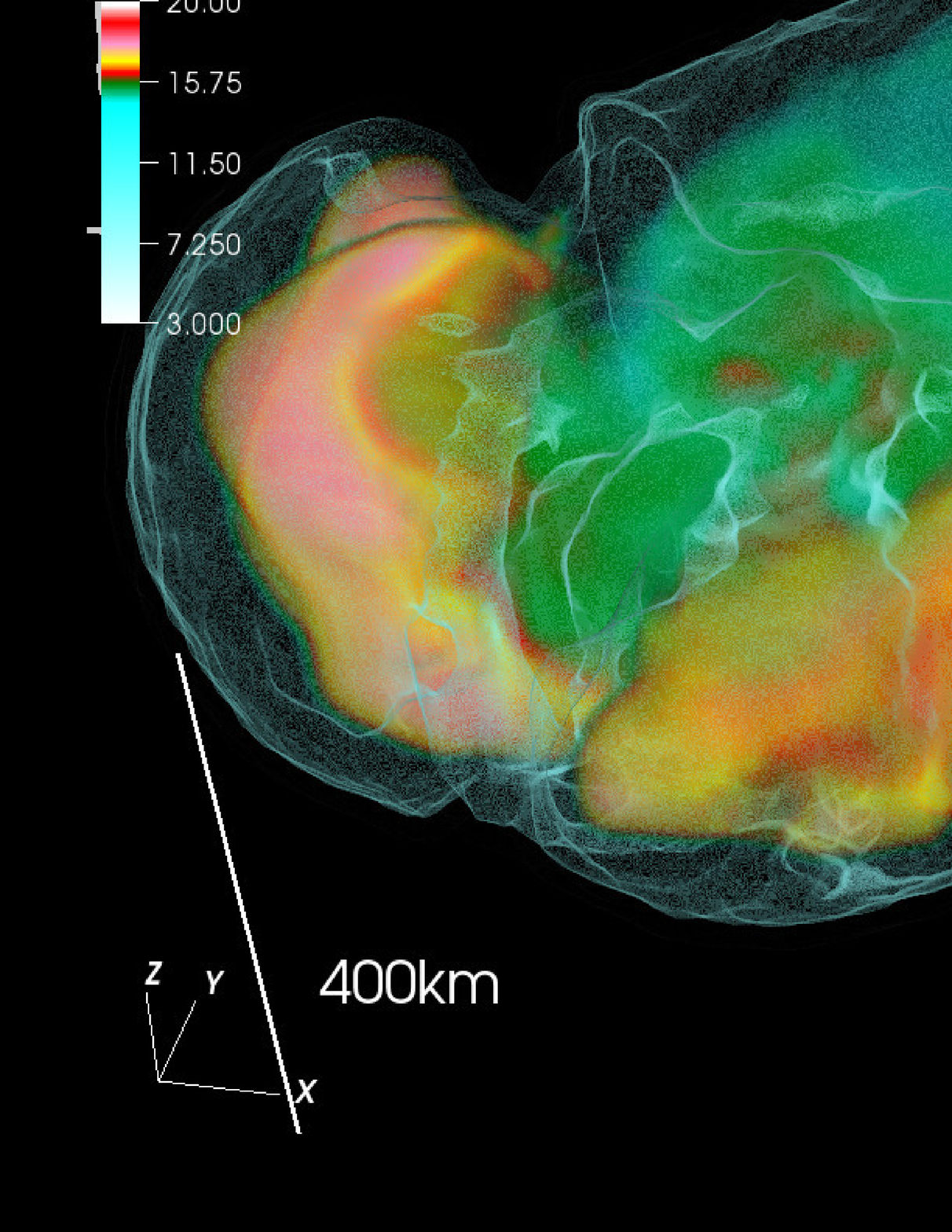}
\includegraphics[width=50mm,angle=0.]{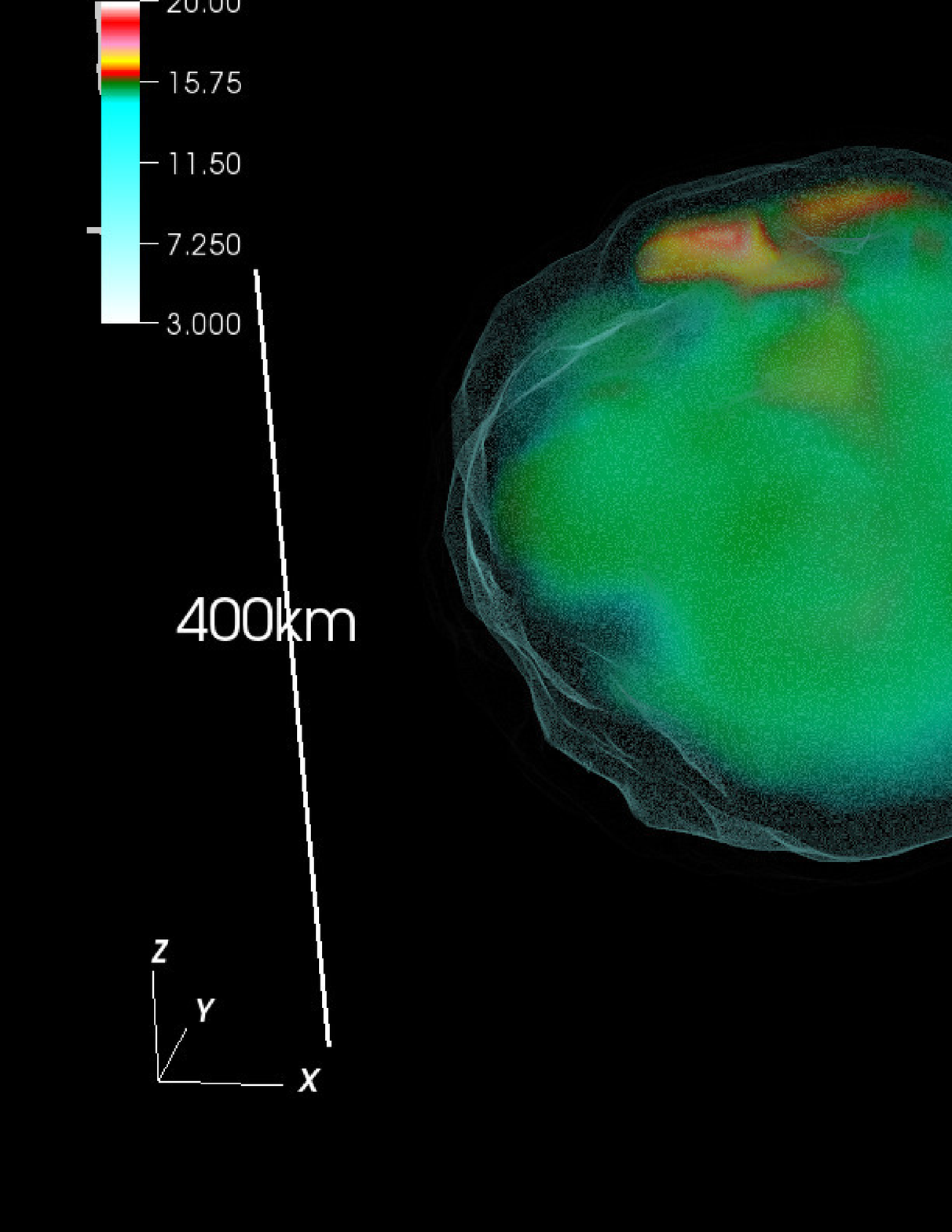}
\includegraphics[width=50mm,angle=0.]{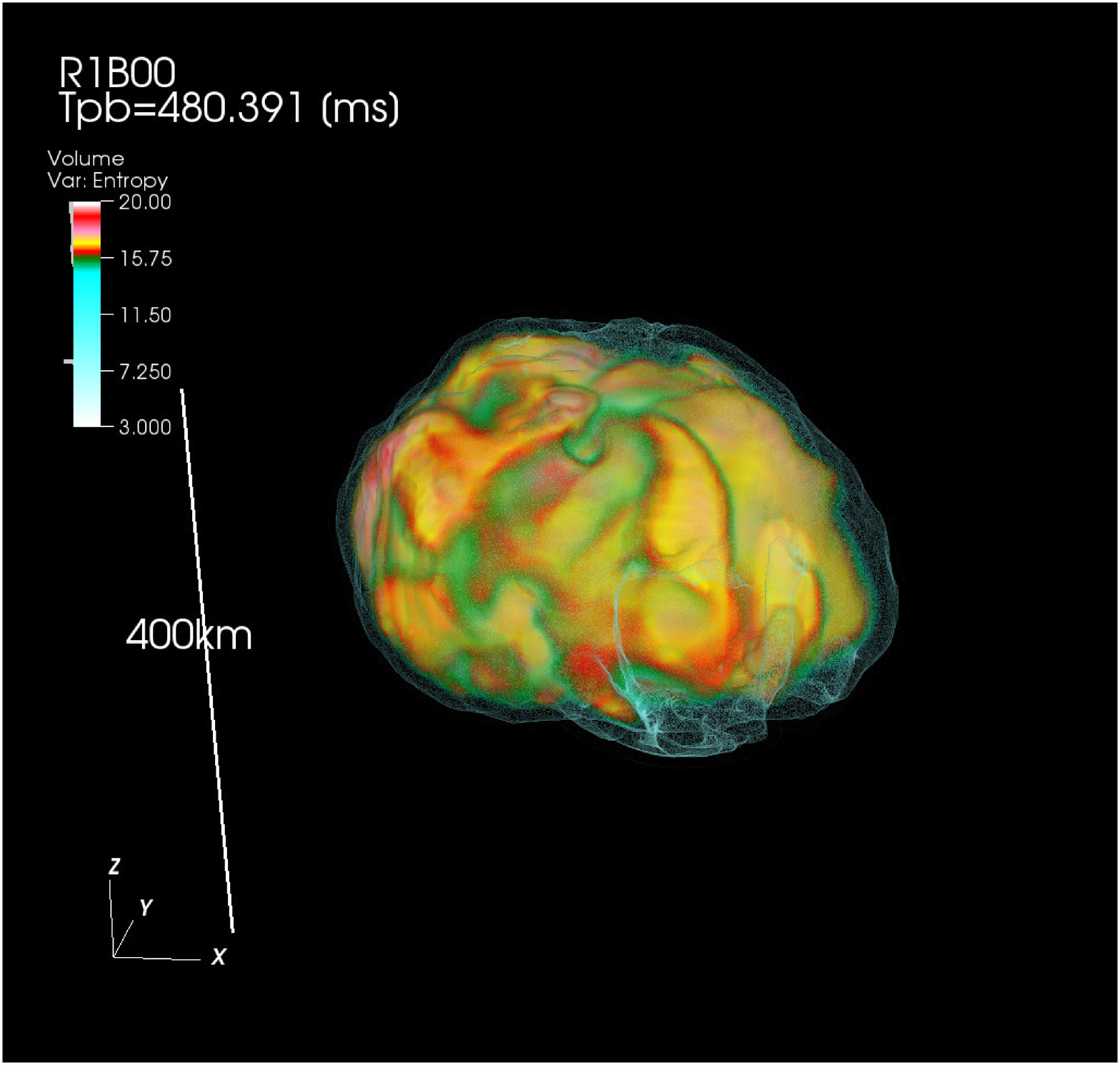}
  \caption{Same as Fig.~\ref{fig:R1B12_Ent_Beta}, but for only entropy of non-magnetized models at different time slices $t_{\rm pb}\sim245$, 370, and 500 ms.
  The upper and lower panels are for model R0B00 and R1B00, respectively.
  Note again that the white line denotes the length scale that is parallel to the rotational axis ($z$-axis).
  \label{fig:R0R1B00_Ent}
}
\end{center}
\end{figure*}
%%%%%%%%%%%%%%%%%%%%%%%%%%%%%%%%%%%%%%
We begin with a brief description of the postbounce evolution of all the three models in this work.
After the start of calculation ($t=0$), the core bounce occurs at $t=0.261$ s, $0.264$ s, and $0.265$ s for model R0B00, R1B00, and R1B12, respectively. The central (maximum) rest mass density $\rho_{\rm max}$ reaches $4.42\times10^{14}$ g cm$^{-3}$ (model R0B00), $4.37\times10^{14}$ g cm$^{-3}$ (R1B00), and $4.35\times10^{14}$ g cm$^{-3}$ (R1B12). A monotonic feature that rapid rotation and high initial magnetic field delay the bounce time and decrease $\rho_{\rm max}$ is due to the stronger centrifugal force and magnetic pressure at bounce.
The lapse function at the center also shows the similar trend, where it takes the smallest and highest value for model R0B00 and R1B12, respectively. For the computed three models, $\rho_{\rm max}$ and the (minimum) lapse function evolve with time after bounce, keeping the above trend at bounce (for example, smallest $\rho_{\rm max}$ for model R1B12 relative to other models).

To visualize the postbounce evolution,  Fig.~\ref{fig:R1B12_Ent_Beta} shows the volume rendered entropy (upper three panels) and inverse of the plasma $\beta$ for model R1B12 in the logarithmic scale (lower three panels) at selected postbounce times ($t_{\rm pb}$). Here the plasma $\beta$ is defined by the ratio of the gas to the magnetic pressure, i.e., $\beta\equiv P_{\rm gas}/P_{\rm mag}$.
After bounce, the formation of the bipolar flow can be clearly seen in the left panels.
Inside the expanding blobs, the magnetic pressure dominates over the gas pressure as shown by the yellowish region ($\log_{10}\beta^{-1}\gtrsim0.5$) in the lower panels. This is a clear evidence of the magnetorotationally-driven shock revival for model R1B12. 
As an important 3D effect, we see that the shock morphology is less collimated compared to the previous 2D axisymmetric studies, although similar initial rotation and magnetic fields were adopted \citep{Burrows07,Takiwaki09,Moesta14,Obergaulinger17}.
The middle panels show that the jet head is not aligned with the rotational axis at $t_{\rm pb}\sim100$ ms, but is displaced from the axis (indicated by the deviation from the white line).
In Sec.~\ref{sec:Non-axisymmetric instabilities inside the MHD outflow}, we will discuss the reason of this in more detail.

Fig.~\ref{fig:R0R1B00_Ent} shows the volume rendered entropy structure for models R0B00 (upper row panels) and R1B00 (lower row panels) from $t_{\rm pb}\sim245$ ms to $\sim500$ ms.
Comparing with the uni-/bipolar like structure seen in the magnetized model R1B12, the shock morphology of these two non-magnetized models is obviously different.
Models R0B00 and R1B00 show roundish and oblate shock morphology, respectively.
During our simulation time up to $t_{\rm pb}\sim500$ ms, we do not find a shock revival in these two non-magnetized models.

\subsection{Shock Wave Evolution}
\label{sec:Shock Wave Evolution}
%%%%%%%%%%%%%%%%%%%%%%%%%%%%%%%%%%%%%%
\begin{figure}[htbp]
\begin{center}
\includegraphics[width=70mm,angle=-90.]{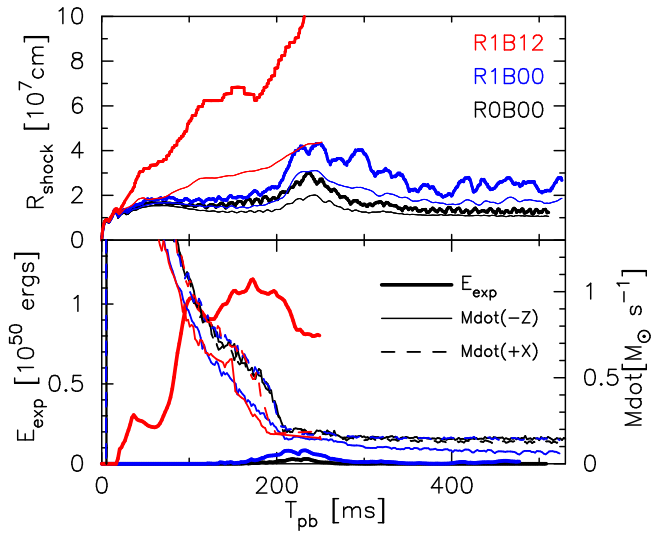}
  \caption{Top: Maximum (thick lines) and averaged (thin) shock radii ($R_{\rm shock}$) are plotted as a function of the postbounce time. Bottom: Time evolution of the diagnostic explosion energy ($E_{\rm exp}$, thick lines) and the mass accretion rate ($\dot M$, thin solid/dashed lines) for all models. In each panel, the color indicates the model name; red (R1B12), blue (R1B00), and black (R0B00).
  For the mass accretion rate, we first measure the mass flux just above the shock surface on the negative $z$- and positive $x$-axes and then multiply them by $4\pi R_{\rm shock}^2$. Here $R_{\rm shock}$ is the corresponding shock position.
  \label{fig:RshockEexp}
}
\end{center}
\end{figure}
%%%%%%%%%%%%%%%%%%%%%%%%%%%%%%%%%%%%%%
Fig.~\ref{fig:RshockEexp} shows the maximum (thick lines) and averaged (thin) shock radii in top panel, the time evolution of the diagnostic explosion energy $E_{\rm exp}$ and mass accretion rate ($\dot M$) in the bottom panel for model R0B00 (black line), R1B00 (blue line), and R1B12 (red line), respectively.
Here, $E_{\rm exp}$ is defined by
\begin{equation}
    E_{\rm exp}=\int_{\tau>0} \sqrt{\gamma}\tau dx^3,
\end{equation}
 which is analogous to Eq.(2) of \cite{BMuller12a}, but takes into account the additional contribution from magnetic fields.
For the mass accretion rate, we first measure the mass flux just above the shock surface on the negative $z$-axis and positive $x$-axis and then multiply them by $4\pi R_{\rm shock}^2$. Here $R_{\rm shock}$ is the corresponding shock position.
Since model R1B12 shows unipolar-like explosion mainly toward negative $z$-axis, we show the value measured on that axis.
The value on positive $x$-axis can be considered as a typical value along the equatorial plane.

From the top panel, one can see that the shock revival is not obtained for the non-magnetized models R0B00 (black line) and R1B00 (blue line) for the simulation time, whereas the shock propagates outwards in the magnetized model R1B12 (red line). The shock is slightly energetized at $t_{\rm pb}\sim180$ ms for model R1B12 and at $t_{\rm pb}\sim200$ ms for models R0B00/R1B00, when the Si/O interface accretes onto the shock. This leads to the runaway shock expansion for model R1B12, whereas it only 
results in the slight shock expansion maximally up to the radius of $\sim 400$ km for model R1B00, gradually shifting to the standing shock later on (see, blue and black lines).
The time when the Si/O interface accretes onto the shock differs about $\sim20$ ms between model R1B12 and the other two models.  The time lag is because of the difference in the (maximum) shock position 
($\sim4\times10^7$ cm)  at $t_{\rm pb}\sim180$ ms. Since the typical accretion velocity is $\sim2\times10^9$ cm s$^{-1}$ there, this can be translated into the crossing time of $\sim20$ ms, which is consistent with the time difference.
The mass accretion rate in Fig.~\ref{fig:RshockEexp} also supports this.
In the lower panel, the mass accretion rate measured on the negative $z$-axis $\dot M(-Z)$ for model R1B12 (red thin solid line) shows the fastest time of accretion of Si/O interface at $T_{\rm pb}\sim182$ ms, while it accretes at $T_{\rm pb}\sim205$ ms in model R0B00 (black thin lines) and also along the equatorial plane in model R1B00 (blue thin dashed line which is overlapped by the black lines).
Therefore, the aforementioned shock expansion can be explained by a sudden reduction of mass accretion rate in association with the accretion of Si/O interface.

The diagnostic explosion energy in the bottom panel basically correlates with the shock evolution.
In the successful explosion model R1B12, the diagnostic explosion energy increases significantly faster than the other two non-explosion models already at $\sim20$ ms after bounce.
It reaches $\sim10^{50}$ erg around $t_{\rm pb}\sim100$ ms.
The value $E_{\rm exp}\sim10^{50}$ erg at the time when the shock reaches $R_{\rm shock}\sim1000$ km is very similar to the ones in previous 2D \citep{Takiwaki09,Obergaulinger17} and 3D \citep{Obergaulinger19} studies with the similar initial rotation and magnetic fields strength.
In the non-magnetized models R0B00 and R1B00, $E_{\rm exp}$ temporally reaches $\sim10^{49}$ erg at $T_{\rm pb}\sim220$ ms when the Si/O interface accretes and a temporal shock expansion occurs, though it soon decreases.

We can also find a typical signature of SASI in the evolution of shock radii.
From top panel in Fig.~\ref{fig:RshockEexp}, a time modulation is visible in the maximum shock radii, particularly in the model R0B00 (thick black line) for $t_{\rm pb}\gtrsim100$ ms.
%%%%%%%%%%%%%%%%%%%%%%%%%%%%%%%%%%%%%%
\begin{figure}[htbp]
\begin{center}
\includegraphics[width=110mm,angle=-90.]{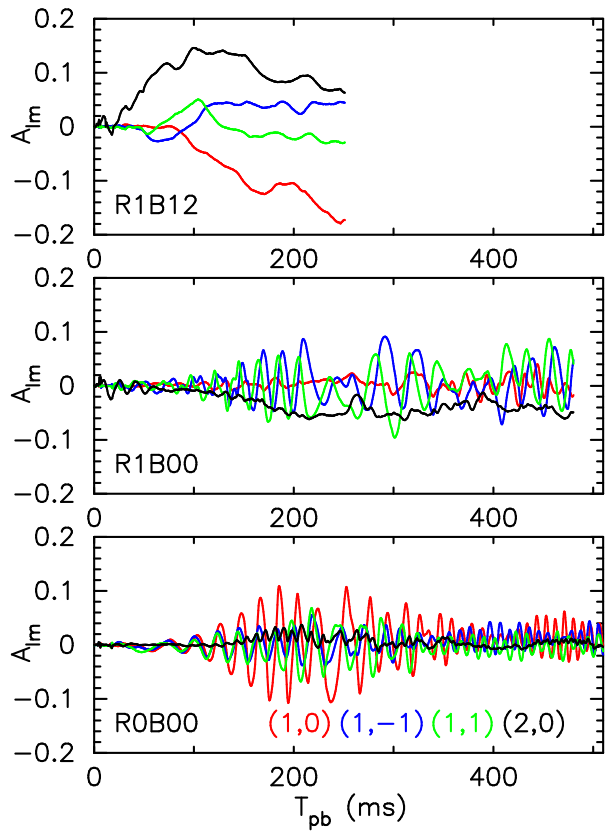}
  \caption{Time evolution of normalized mode amplitudes $A_{\ell m}$ of spherical polar expansion of the shock surface $R_{\rm shock}(\theta,\phi)$.
   The top, middle, and bottom panel is for model R1B12, R1B00, and R0B00, respectively.
  Note that we plot only several dominant modes, $(\ell,m)=(1,0)$, $(1,\pm1)$, and $(2,0)$, denoted in the bottom panel.
}  \label{fig:Alm}
\end{center}
\end{figure}
%%%%%%%%%%%%%%%%%%%%%%%%%%%%%%%%%%%%%%
Such a modulation reflects the appearance of SASI \citep{Scheck06,Foglizzo06}.
To see more quantitatively the shock morphology and also the dominant SASI mode, we plot time evolution of normalized mode amplitudes $A_{\ell m}\equiv c_{\ell m}/c_{00}$ of spherical polar expansion of the shock surface $R_{\rm shock}(\theta,\phi)$ for several dominant modes in Fig.~\ref{fig:Alm}.
Here we adopt the same definition for $c_{\ell m}$ as in \cite{burrows12} with $\ell$ and $m$ representing the quantum number with respect to the real spherical harmonics of $Y_\ell^m$, respectively.

In the top panel, the dominant mode is $(\ell,m)=(2,0)$ (black line) for the first $\sim120$ ms after bounce.
Since its sign is positive, the shock morphology is prolate as also shown in the left and center columns in Fig.~\ref{fig:R1B12_Ent_Beta}.
However, for $t_{\rm pb}\gtrsim120$ ms in the same model R1B12, $(\ell,m)=(1,0)$ (red line) gradually takes over as the dominant term with its sign being negative.
Therefore the shock morphology at the end of simulation time is unipolar toward the negative $z$-axis, which is again consistent with the right column in Fig.~\ \ref{fig:R1B12_Ent_Beta}.
In the middle panel, R1B00 shows that $A_{20}$  becomes negative for $t_{\rm pb}\gtrsim50$ ms which reflects a rotating oblate spheroid (see, bottom panels in Fig.~\ref{fig:R0R1B00_Ent}).
At the same time, $(\ell,m)=(1,\pm1)$ (blue and green lines) also show comparable amplitudes with that of (2,0), but with clear quasi-periodic oscillations.
Between these two $|m|=1$ modes, i.e., $(\ell,m)=(1,1)$ and $(1,-1)$, a phase shift seemingly with $\sim\pi/2$ exists which indicates that the spiral SASI motion appears \citep{Blondin07_nat}.
In the non-rotating model R0B00, all the three modes with $\ell=1$ and $m=0,\pm1$ show basically the same amplitude with almost no phase shift up to $t_{\rm pb}\sim120$ ms.
Therefore, the dominant SASI mode is the sloshing mode firstly after bounce.
Afterward the $(1,0)$ mode gradually decouples from the other two different azimuthal modes.
There seems to be a phase shift of $\sim\pi/2$ between (1,0) (red line) and the other two with $(1,\pm1)$ (green and blue).
This can be explained by the dominant SASI motion changing from the sloshing motion to the spiral one around $t_{\rm pb}\sim120$ ms. Note that the growth of the spiral SASI in the non-rotating progenitors \citep{Blondin07_nat} is consistent with the outcomes of previous 3D core-collapse models \citep{Hanke13,KurodaT16ApJL,Ott18}.

\subsection{Non-axisymmetric instabilities inside the MHD outflow}
\label{sec:Non-axisymmetric instabilities inside the MHD outflow}

In this subsection, we discuss non-axisymmetric instabilities inside the MHD outflow and their potential impact on the shock evolution.
In a 3D-GR model using the similar precollapse rotation rate and magnetic fields to our model R1B12, \citet{Moesta14} observed the appearance of the kink instability \citep{Lyubarskii99,Begelman98,Narayan09}.
According to their analysis, the linear growth of the kink instability shortly starts after bounce, which is followed by the non-linear phase already at $t_{\rm pb}\sim20$ ms.
At that moment, the jet barycenter showed a significant displacement from the rotational axis, which is one of the main features of the growth of the kink instability, leading to a broader and less energetic outflow compared to the counterpart axisymmetric case.
We also check if this instability appears and affects the dynamics of outflow in model R1B12.

The condition $|b_{\phi}/b_z| > \varpi/L$, i.e., the well-known Kruskal-Shafranov criterion, is the major factor that determines whether the system is unstable to the most dominant screw mode, i.e., for $|m|=1$ mode with a condition $mb_\phi<0$.
Here $L$ and $\varpi$ denote the inverse of minimal wave number of the unstable mode propagating parallel to the rotational axis and distance from the rotational axis, respectively.
In a sufficiently rapidly rotating case, one should also take into account the rotational stabilizing effect that relaxes the Kruskal-Shafranov criterion to $|b_{\phi}/b_z| > \Omega\varpi$ \citep{Tomimatsu01}, where $\Omega$ is the angular frequency in geometric units.
In our magnetized model R1B12, the toroidal magnetic field dominates over the poloidal one $|b_{\phi}/b_z|>1$ just above the PNS core ($z\sim10-50$ km).
Such a configuration is usually seen in the magnetized collapse model as the initial poloidal field can be very efficiently converted into the toroidal one mainly through the field-wrapping.
As a consequence, the value $|b_{\phi}/b_z|/(\Omega\varpi)$ inside the MHD outflow reaches $\mathcal{O}(10^2\sim10^3)$ in our model.
We therefore consider that the MHD outflow appeared in our model R1B12 can also be subject to the kink instability.

Following \cite{Moesta14}, we monitor how the barycenter of MHD outflow is displaced from the rotational/magnetic field axis, i.e., $z$-axis.
We take the same definition for the barycenter $x^i_{\rm c}$ written by \citep{Moesta14}
\begin{eqnarray}
x^i_{\rm c}(z)=\frac{\int ds\,x^i\,P_{\rm mag}}{\int ds\, P_{\rm mag}},
\end{eqnarray}
for $i=1$ and $2$, where we perform the surface integral $\int ds$ over the domain with $|x,y|\le50$ km at $z=\pm50$ km.
In addition, to see the mode propagation direction properly in a rotating system, we map the original Cartesian coordinates $x^i$ to a rotating frame $\bar{x}^i$ by
\begin{eqnarray}
\bar{x}^i=Q^i_jx^j
\end{eqnarray}
with $Q^i_j$ being the usual rotation operator with respect to $z$-axis
\begin{equation}
Q^i_j=\left[
\begin{array}{ccc}
\cos{\Theta} & \sin{\Theta} & 0 \\
-\sin{\Theta} & \cos{\Theta} & 0 \\
0 & 0 & 1 \\
\end{array}
\right].
\end{equation}
$\Theta(t,z)$ measures the cumulative rotation angle of the system at a given slice $z(=\bar{z})$ after core bounce and is given by a following rough estimation
\begin{eqnarray}
\label{eq:Theta}
\Theta(t,z)=\int_{t_{\rm{cb}}}^t dt' \,\bar{\omega},
\end{eqnarray}
where $\bar{\omega}(t,z)$ is the mean angular frequency of the plane.
Since the PNS differentially rotates, the rotational angle $\Theta(t,z)$ is just a rough measurement.
We evaluate the mean angular frequency $\bar{\omega}(t,z)$ simply by
\begin{eqnarray}
\bar{\omega}(t,z)=\frac{\int ds\, \omega^z\, \rho}{\int ds\,\rho},
\end{eqnarray}
where $\omega^z=v^z/\sqrt{x^2+y^2}$ is the angular frequency measured in the Eulerian frame and we use the rest mass density as a weight.
After mapping, we plot the barycenter $x^i_{\rm c}$ on the rotating plane $\bar{x}\bar{y}$ at $\bar{z}=\pm50$ km.
%%%%%%%%%%%%%%%%%%%%%%%%%%%%%%%%%%%%%%
\begin{figure}[htbp]
\begin{center}
\includegraphics[width=140mm,angle=-90.]{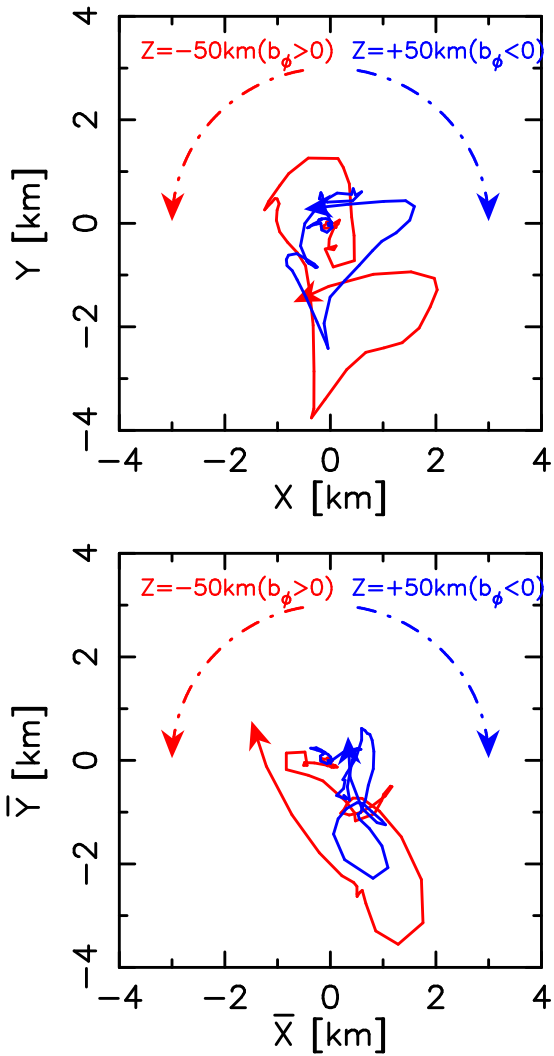}
  \caption{Solid lines: Trajectories of the barycenter of MHD outflow on the original $xy$ plane (top) and on the rotating $\bar{x}\bar{y}$ plane (bottom). The color represents the position of the planes either at $z=50$ km (blue lines) or $z=-50$ km (red).
  Time evolution is indicated by the arrow.
  Dash-dotted lines: We show direction of $b_\phi$, averaged over $\varpi\lesssim40$ km, which is clockwise ($b_\phi<0$) and counterclockwise ($b_\phi>0$) for $z>0$ and $z<0$, respectively, on these planes.
  \label{fig:KINK}
}
\end{center}
\end{figure}
%%%%%%%%%%%%%%%%%%%%%%%%%%%%%%%%%%%%%%

In top and bottom panels of Fig.~\ref{fig:KINK}, we show the trajectory of barycenter of MHD outflow (solid lines) on the original $xy$ and rotating $\bar{x}\bar{y}$ planes at $z=\pm50$ km.
To highlight the initial linear growth phase, we show only from the bounce time up to $t_{\rm pb}=30$ ms that is indicated by the arrow.
In addition, we show direction of $b_\phi$ averaged over $\varpi\lesssim40$ km by dash-dotted line for reference.
Because of our initial purely poloidal magnetic field with dipole-like structure orienting toward positive $z$-axis, direction of the toroidal component generated after core-collapse mainly through the field wrapping is basically clockwise ($b_\phi<0$) and counterclockwise ($b_\phi>0$) for $z>0$ and $z<0$, respectively, on these planes.
Note that the positive $z$-axis points toward us and from the condition $mb_\phi<0$ that selects the leading mode to develop, the propagation direction of the most unstable mode in a comoving frame is expected to be counterclockwise ($m=1$) and clockwise ($m=-1$) for $z>0$ and $z<0$, respectively.

From top panel in Fig.~\ref{fig:KINK}, both of the solid lines show basically counterclockwise propagation direction, i.e., $m=1$ mode.
In top panel, the mode propagation direction (blue solid arrow) is counterclockwise (i.e., $m=1$) and is opposite to that of $b_\phi(<0)$ (blue dash-dotted arrow), for the region with $z>0$, meaning that it is consistent with a linear analysis $mb_\phi<0$.
On the other hand, in the same top panel, both solid and dash-dotted red arrows are pointing toward the same counterclockwise direction on the plane at $z=-50$ km, which is not in accordance with the theoretical expectation $mb_\phi<0$.
We think that this inconsistency seen in red arrows (top panel) is apparent as, from bottom panel, the red solid arrow in the rotating frame is showing a clockwise propagation direction (i.e., $m=-1$) opposite to that of $b_\phi(>0)$.
These facts support that the kink instability likely appears, displaces the shock center, and consequently makes the shock morphology broader compared to the corresponding 2D model.

We should also mention another relevant non-axisymmetric instability that might influence on the growth of the above kink instability.
As we have already mentioned, the ratio of rotational to gravitational potential energy after bounce in both of our rotating models reaches several percent, which makes the PNS core being subject to the low-$T/W$ instability \citep{Watts05,Saijo06}.
Once the instability appears, it produces an instability mode that propagates in the same direction as the fluid motion, i.e., this time with $m=1$ mode in both the northern and southern hemispheres.
Therefore, it means that the two different instabilities, namely the low-$T/W$ and kink instabilities, could simultaneously exist possibly with the same $m=+1$ mode for $z>0$ and with two opposite $m=\pm1$ modes for $z<0$ breaking the parity between northern and southern hemispheres.

It is beyond the scope of this paper to quantify how the two instabilities coexist, how they affect the PNS core dynamics and the disrution of the bipolar flows as seen in model R1B12.
%however, if they were dynamically relevant to the MHD outflow, it could be a possible reason for the disruption of bipolar structure seen in model R1B12.
%Without solid evidence, We speculate that relatively weak explosion ($E_{\rm exp}\sim10^{50}$ erg from Fig.~\ref{fig:RshockEexp}) could be a 
%possible outcome of the kink instability
%and low-$T/W$ instabilities leading to a weaker explosion in one of bipolar outflows.
Once the bipolar flows are disrupted, the mass accretion rate becomes higher on the weaker explosion side as a consequence of deflection of mass accretion on the stronger explosion side. This could explain a relatively weak explosion ($E_{\rm exp}\sim10^{50}$ erg) of model R1B12.
%f  the weakening and sometime results in a disruption of outflow on this side.
Apparently we need more studies with varying the initial magnetic fields and rotational profiles systematically in order to clarify the disruption mechanism of the MHD outflows.

%%%%%%%%%%%%%%%%%%%%%%%%%%%%%%%%%%%%%%
\begin{figure*}[htbp]
\begin{center}
\includegraphics[width=80mm,angle=-90.]{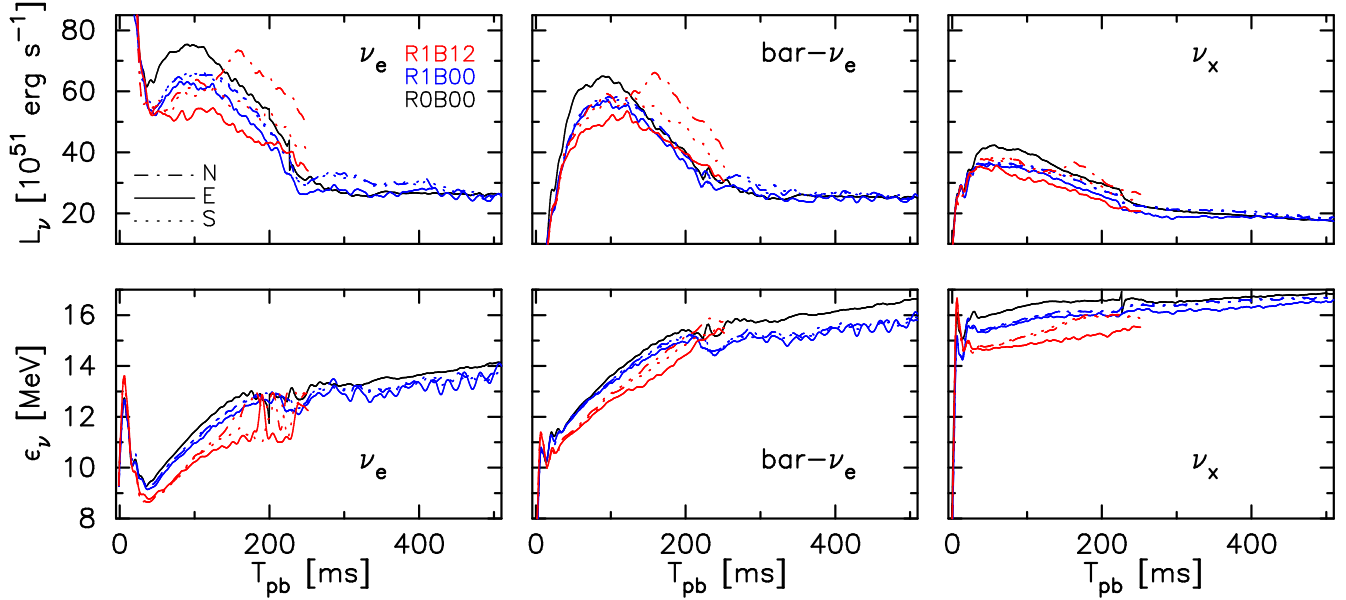}
  \caption{We plot (viewing-angle-dependent) neutrino luminosity $L_\nu$ (top row) and mean neutrino energy $\varepsilon_\nu$ (bottom row) at specific observer angles for $\nu_e$ (left), $\bar\nu_e$ (middle), and $\nu_x$ neutrinos (right). These quantities are estimated at a radius of $r=400$ km.
  We choose three observer angles denoted by N (north pole), E (equatorial plane, here represented by positive $x$-axis), and S (south pole).
  For the non-rotating model R0B00, we plot only spherical averaged values (solid black line) for simplicity.
  \label{fig:LnuEnu}
}
\end{center}
\end{figure*}
%%%%%%%%%%%%%%%%%%%%%%%%%%%%%%%%%%%%%%
\subsection{Rotational Effects on Neutrino Profiles}
\label{sec:Rotational Effects on Neutrino Profiles}
The time modulation of CCSN neutrino signals reflects the hydrodynamics evolution of the postbounce core (e.g., \citet{Tamborra13,walk19}, and \citet{mirizzi16} for a review). 
In this section, we describe how we can make the connection between the core dynamics and neutrino signals.
In Fig.~\ref{fig:LnuEnu}, we plot the neutrino luminosity $L_\nu$ (top row) and mean neutrino energy $\varepsilon_\nu$ (bottom row) for specific observer angles for electron type (left), anti-electron type (middle), and heavy lepton type neutrinos (right).
Here we evaluate these signals by averaging the neutrino's energy flux at $r=400$ km following \cite{Tamborra14ApJ}.
We choose three observer angles relative to the rotational axis that are denoted by N (north pole), E (equatorial plane, here represented by the positive $x$-axis), and S (south pole).
To prevent too many lines, we plot only spherical averaged values for the non-rotating model R0B00 (solid black line) as it shows basically no significant asymmetry.

Common features among all models are as follows.
The neutrino luminosities of all flavours plateau at $t_{\rm pb}\sim50-100$ ms.
At that moment, the luminosities reach $L_\nu\sim6\times10^{52}$ erg s$^{-1}$ for $\nu_e$ and $\bar\nu_e$ and $L_\nu\sim3.5\times10^{52}$ erg s$^{-1}$ for $\nu_x$.
Although such values depend on the progenitor star, EOS, and neutrino matter interactions employed, the peak luminosities are in good agreement with those in recent studies with detailed neutrino transport \citep{BMuller17,O'Connor18,Summa18,Vartanyan19}.
The luminosities become nearly constant at $t_{\rm pb}\sim220$ ms when the mass accretion decreases suddenly.
We can also see how the progenitor rotation and magnetic field affects the neutrino signals.
The non-rotating model R0B00 shows basically the highest luminosity and mean energy in all flavor of neutrinos (see black lines).
Meanwhile, the rotating magnetized model R1B12, which explodes shortly after bounce, shows lowest values in both luminosities and mean energies, though there is a slight observer angle dependence.
The model R1B00 appears in between them.
Such features stem from that the most compact PNS of R0B00 without being subject to the rotational flattening emits higher neutrino luminosities and energies due to its hotter core temperature.
On the other hand, the rotating magnetized model R1B12, which shows a lower mass accretion rate due to the centrifugal force and also experiences the mass ejection through bipolar outflow, has a less compact PNS leading to lower neutrino energies and luminosities.

%%%%%%%%%%%%%%%%%%%%%%%%%%%%%%%%%%%%%%
\begin{figure*}[htbp]
\begin{center}
\includegraphics[width=70mm,angle=-90.]{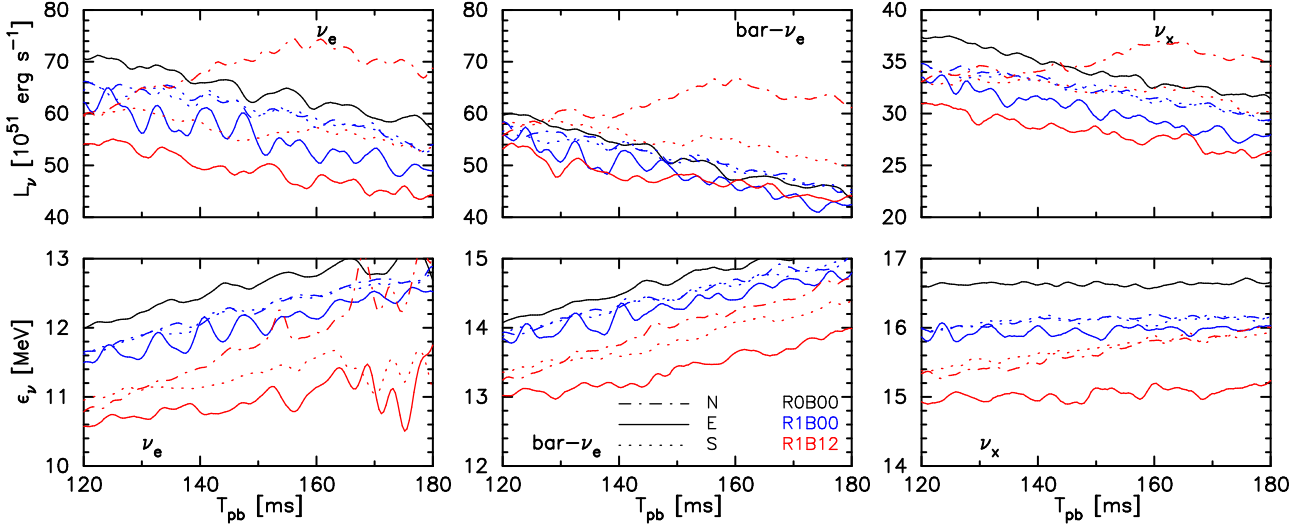}
  \caption{Same as Fig.~\ref{fig:LnuEnu}, but for the postbounce time from $t_{\rm pb}=120$ ms to 180 ms.
  \label{fig:LnuEnuMini}
}
\end{center}
\end{figure*}
%%%%%%%%%%%%%%%%%%%%%%%%%%%%%%%%%%%%%%
Next, we focus on the viewing angle dependence of the neutrino signals. 
In Fig.~\ref{fig:LnuEnuMini}, we show a magnified view of Fig.~\ref{fig:LnuEnu} from $t_{\rm pb}=120$ ms to 180 ms.
In both the rotating models R1B00 and R1B12, the neutrino luminosity and energy observed along the equatorial plane (solid red and blue lines) show the lowest value compared to those along the rotational axis (dash-dotted and dotted lines labeled by N and S). This is because of the rotational flattening of the PNS, 
where the neutrino-sphere radius along the equatorial plane is located outward than that of the rotational axis, making the neutrino temperature seen along the equator lower than that from the rotational axis \citep[e.g.,][]{Kotake03,Ott08,harada19}.
%%%%%%%%%%%%%%%%%%%%%%%%%%%%%%%%%%%%%%
\begin{figure}[htbp]
\begin{center}
\includegraphics[width=100mm,angle=-90.]{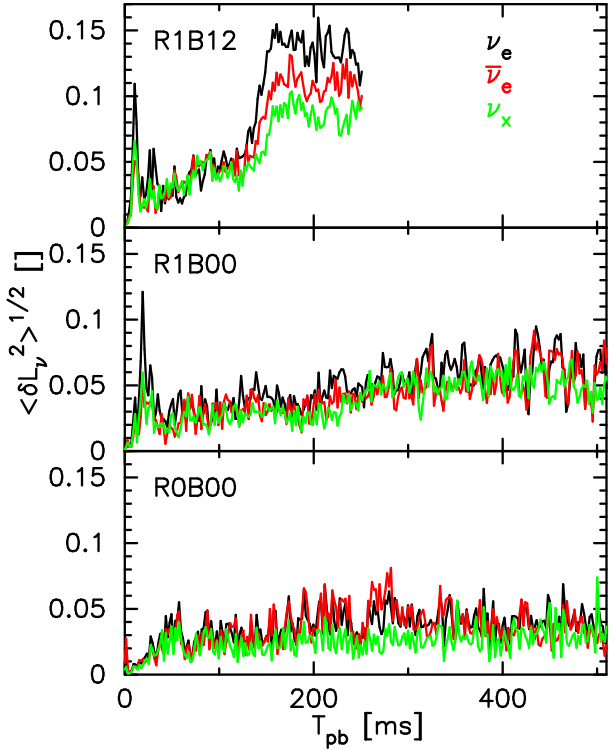}
  \caption{Time evolution of the root mean square variation $\sqrt{\langle\delta L_\nu^2\rangle}$ around the angle averaged neutrino luminosities for all neutrino flavours.
  From top, we show the value in model R1B12, R1B00, and R0B00.
  \label{fig:Lnu_RMS}
}
\end{center}
\end{figure}
%%%%%%%%%%%%%%%%%%%%%%%%%%%%%%%%%%%%%%
To show the viewing angle dependence more quantitatively, we plot the root mean square (RMS) variation $\sqrt{\langle\delta L_\nu^2\rangle}$ around the angle averaged neutrino luminosities $\langle L_\nu\rangle$ in Fig. \ref{fig:Lnu_RMS}, where $\sqrt{\langle\delta L_\nu^2\rangle}$ is defined by
\begin{eqnarray}
\sqrt{\langle\delta L_\nu^2\rangle}\equiv \sqrt{\frac{1}{4\pi}
\int_{R=400{\rm km}} ds \left(\frac{L_\nu-\langle L_\nu\rangle}{\langle L_\nu \rangle} \right)^2
}.
\end{eqnarray}
As we have mentioned, the rotational flattening of the PNS produces larger viewing angle dependence that is clearly seen by larger RMS values in R1B00 than those in R0B00.
Furthermore, model R1B12 shows the largest variance due to its highly aspherical explosion morphology.
Another remarkable feature is that there is a hierarchy by neutrino species of $\nu_e>\bar\nu_e>\nu_x$, which is most significant in model R1B12 and is diminished in non-rotating non-exploding model R0B00.
We note that the hierarchy is different from previous report $\bar\nu_e>\nu_e>\nu_x$ by \cite{Vartanyan19b}.
Although we do not know the exact reason of the difference, the hierarchy basically indicates how large each of the neutrino spheres deforms and, thus, may depend on both the adopted neutrino opacities and transport method.

There is yet another neutrino signature for model R1B00. Seen from the equatorial plane (blue solid line),
  a clear periodic time modulation can be seen.
On the other hand, the modulation is hard to be seen from the rotational axis (blue dash-dotted and dotted lines).
Furthermore, the degree of the rotational effect differs depending on the neutrino flavour.
It is particularly strong in $\nu_e$ and becomes weaker in order of $\bar\nu_e$ and $\nu_x$.
 Fig.~\ref{fig:LnuEnuMini} shows that $\nu_e$ signals have a time modulation with amplitudes of $\sim5\times10^{51}$ erg s$^{-1}$ and $\sim0.5$ MeV for the luminosity and mean energy, respectively, while those values decrease to $\sim1\times10^{51}$ erg s$^{-1}$ and $\sim0.2$ MeV for $\nu_x$.
Such a modulation was first discussed in \cite{Takiwaki18} and is associated with the growth of the so-called low-$T/W$ instability \citep{Ott05,Saijo06,Watts05} and the (one-armed) spiral flows.
In both of our rotating models, the ratio of rotational to gravitational potential energy after bounce reaches several percent, which is close to the onset of the low-$T/W$ instability. The neutrino spheres of all flavors are located above the PNS core surface at $R\sim10$ km, where the low-$T/W$ instability starts to (typically) develop, and also below the shock which is deformed by the spiral SASI (for model R1B00). Once the two instabilities appear, they can deform the neutrino spheres and potentially be the origin of the neutrino
 time modulation (see \citet{Kazeroni2017} for the possible connection of the two instabilities). However, we note that the smaller modulation in the $\nu_x$ signals seem to favor that the outermost $\nu_e$ sphere is more strongly affected by the spiral SASI.

%%%%%%%%%%%%%%%%%%%%%%%%%%%%%%%%%%%%%%
\begin{figure}[htbp]
\begin{center}
\includegraphics[width=50mm,angle=-90.]{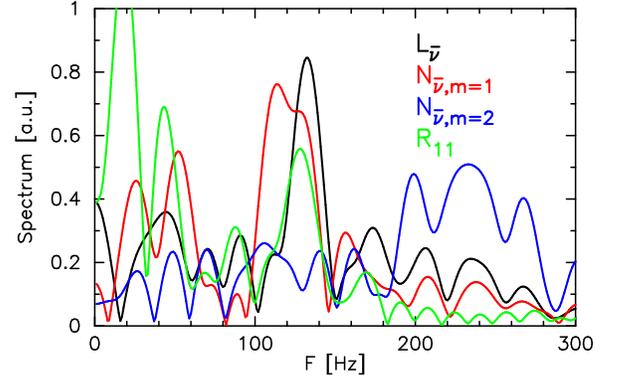}
  \caption{We plot spectra of the (viewing-angle-dependent) neutrino luminosity $L_{\bar\nu_e}$ (black line) corresponding to the blue solid line in the upper middle panel of Fig.~\ref{fig:LnuEnuMini}, of normalized mode amplitudes of the number luminosity $N_{\bar\nu_e,m}$ with the lower index $m$ being the azimuthal mode either $m=1$ (red) or $m=2$ (blue), and of the normalized mode amplitude of spherical polar expansion of the isodensity surface $R_{11}$, extracted at the rest mass density of $\rho=10^{11}$ g cm$^{-3}$, for $(\ell,m)=(1,1)$ (green).
  The vertical axis is in arbitrary unit.
  Here $R_{11}$ is roughly representing the neutrino sphere.
  The spectra are obtained by the Fourier transformation for the time interval of $120\le t_{\rm pb}\le180$ ms.
  \label{fig:LnuDiQdSpec}
}
\end{center}
\end{figure}
%%%%%%%%%%%%%%%%%%%%%%%%%%%%%%%%%%%%%%
Indeed there is a quantitative evidence that the deformation of neutrino sphere creates the time modulated neutrino signals.
In Fig.~\ref{fig:LnuDiQdSpec}, we plot spectra of the (angle-dependent) neutrino luminosity $L_{\bar\nu_e}$ corresponding to the blue solid line in the upper middle panel of Fig.~\ref{fig:LnuEnuMini}, of normalized mode amplitudes of the number luminosity $N_{\bar\nu_e,m}$ for $m=1,2$, and of the normalized mode amplitude of spherical polar expansion of the isodensity surface $R_{11}$ for mode $(\ell,m)=(1,1)$.
Here, $N_{\bar\nu_e,m}$ is evaluated by
\begin{equation}
    N_{\bar\nu_e,m}=\frac{|\int d\phi N_{\bar\nu_e} e^{im\phi}|}{\int d\phi N_{\bar\nu_e}}
\end{equation}
at $R=400$ km and $\theta=90^\circ$ with $N_{\nu}$ being the number luminosity estimated in the same way as the luminosity $L_\nu$ \citep{Tamborra14ApJ}.
Although we here use the number luminosity $N_{\nu}$, we can do the same discussion using the luminosity $L_\nu$.
$R_{11}$ is the isodensity surface extracted at the rest mass density of $\rho=10^{11}$ g cm$^{-3}$ corresponding roughly to the neutrino sphere.
The normalized mode amplitude of spherical polar expansion of $R_{11}$ is evaluated exactly in the same manner as that of the shock surface.
%The spectra are obtained by the Fourier transformation for the time interval of to focus on the time modulation seen in Fig.~\ref{fig:LnuEnuMini}. 
Here we focus on the $\bar\nu_e$ signals  ($120\le t_{\rm pb}\le180$ ms) bearing in mind the detectability \citep{Hyper-K2016,IceCube11} (which will be reported elsewhere).

The black line in Fig.~\ref{fig:LnuDiQdSpec} shows that the time modulation seen in Fig.~\ref{fig:LnuEnuMini} peaks at $F\sim125$ Hz.
This component is mainly composed of $m=1$ neutrino number-flux as the two peaks of red and black lines are appearing nearly the same frequency.
The peak of $N_{\bar\nu_e}$ with $m=2$, which is a daughter mode of $m=1$, appears closely at a double frequency $F\sim240$ Hz of that of $m=1$ as expected, but the $m=2$ mode seems to contribute less to the total neutrino signals than the $m=1$ mode.
Finally, as it is obvious from the peak at $F\sim125$ Hz in green line, the origin of these time modulations of the neutrino signals is $m=1$ deformation of neutrino sphere represented by $R_{11}$.
We thus conclude that the strong spiral SASI appearing in R1B00 deforms the neutrino sphere with the same $m=1$ mode and leads to the characteristic neutrino signals.
%due to the lighthouse effect \citep{Takiwaki18}.

We also mention that we observe a clear north-south asymmetry in neutrino signals in model R1B12 for $t_{\rm pb}\gtrsim120$ ms, i.e., between dash-dotted and dotted red lines, which cannot be seen in the corresponding lines of R1B00.
In this model R1B12, the neutrino emission toward the north pole is significantly stronger than the one toward south.
The excess toward north is consistent with the one-sided explosion to the south pole (see the red line in the top panel of Fig.~\ref{fig:Alm} for $A_{10}$ mode).
Due to the shock expansion mainly toward the south, the mass accretion is stronger in the northern hemisphere, which results in higher accretion luminosities and neutrino energies in the north pole.

\subsection{The Role of Neutrino Heating}
\label{sec:The Role of Neutrino Heating}

%%%%%%%%%%%%%%%%%%%%%%%%%%%%%%%%%%%%%%
\begin{figure*}[htbp]
\begin{center}
\includegraphics[width=75mm,angle=-90.]{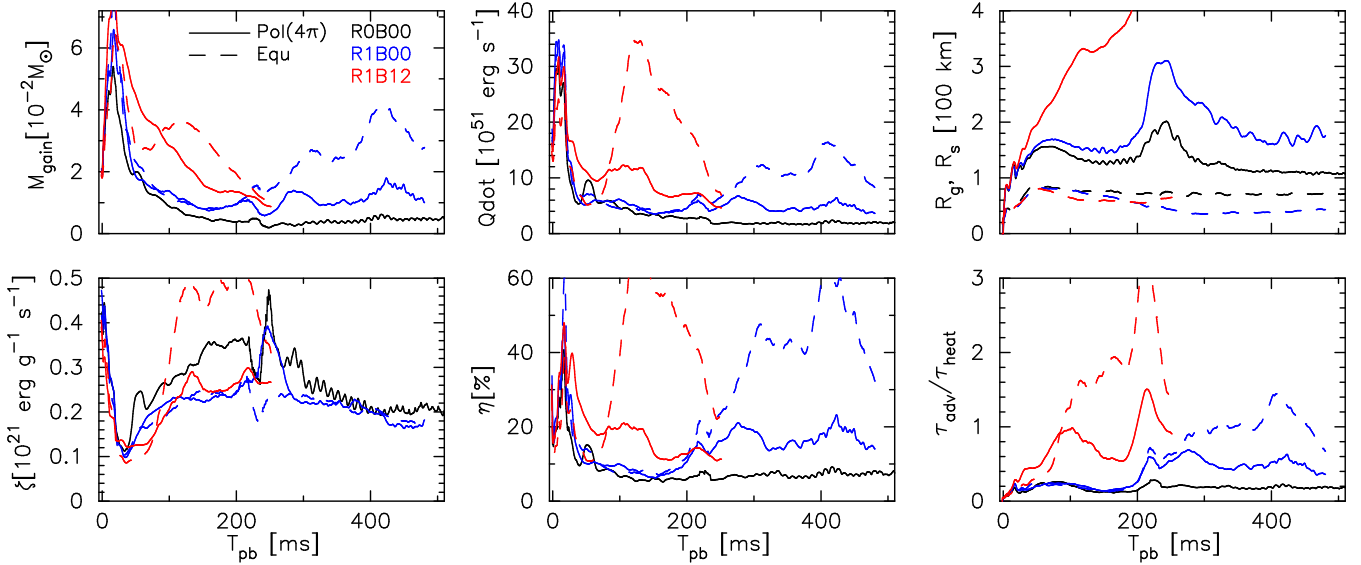}
  \caption{We plot $M_{\rm gain}$ (top left), $\dot Q$ (top middle), $R_{\rm gain}$ and $R_{\rm shock}$ (top right), $\zeta$ (bottom left), $\eta$ (bottom middle), and $\tau_{\rm adv}/\tau_{\rm heat}$ (bottom right) for each model. See text for their definitions.
  The solid and dashed lines represent that the volume/surface integral is performed around the polar axis (labeled by ``Pol'') and equatorial plane (labeled by ``Equ''), respectively.
  Regarding the non-rotating model R0B00, we integrate over all solid angles (solid black line labeled by ``$4\pi$'').
  Note that we show half values for extensive variables, i.e., $M_{\rm gain}$ and $\dot Q$, for model R0B00 for comparison with other models.
  \label{fig:TadvTheat}
}
\end{center}
\end{figure*}
%%%%%%%%%%%%%%%%%%%%%%%%%%%%%%%%%%%%%%
Next we make a comparison of the energetics and discuss the role of neutrino heating among the models, particularly how the neutrinos contribute to the shock expansion.
In Fig.~\ref{fig:TadvTheat}, we plot the mass in the gain region $M_{\rm gain}$ (top left), heating rate $\dot Q$ (top middle), gain and shock radii, $R_{\rm gain}$ and $R_{\rm shock}$, respectively (top right), specific heating rate $\zeta=\dot Q/M_{\rm gain}$ (bottom left), heating efficiency $\eta=\dot Q/(L_{\nu_e}+L_{\bar \nu_e})$ which measures how much of the emergent $\nu_e$ and $\bar\nu_e$'s contribute to the matter heating (bottom middle), and ratio of advection to heating time scale $\tau_{\rm adv}/\tau_{\rm heat}$ (bottom right) for each model.
To obtain these values, we first define the gain radius $R_{\rm gain}(\theta,\phi)$ at each radial direction $(\theta,\phi)$.
$R_{\rm gain}$ is defined at the first point where the net energy deposition rate $\dot q$ becomes zero behind the shock, with $\dot q$ being defined by
\begin{equation}
\dot q\equiv\alpha\sqrt{\gamma}\int d\varepsilon \sum_\nu S_{(\nu,\varepsilon)}^\mu n_\mu.
\end{equation}
Then each value is defined by
\begin{eqnarray}
\label{eq:Mgain}
    M_{\rm gain}&=&\int_{R_{\rm gain}(\theta,\phi)<r<R_{\rm shock}(\theta,\phi)} \rho^\ast dx^3, \\
\label{eq:Qdot}
    \dot Q&=&\int_{R_{\rm gain}(\theta,\phi)<r<R_{\rm shock}(\theta,\phi)} \dot q dx^3, \\
\label{eq:tau_adv}
\tau_{\rm adv}&=&\frac{M_{\rm gain}}{|\int_{r=R_{\rm shock}(\theta,\phi)}\rho^\ast v^r ds|},
\end{eqnarray}
and
\begin{eqnarray}
\label{eq:tau_heat}
    \tau_{\rm heat}&=&\frac{\int_{R_{\rm gain}(\theta,\phi)<r<R_{\rm shock}(\theta,\phi)} \sqrt{\gamma}\tau dx^3}{\dot Q},
\end{eqnarray}
where the surface integral $\int ds$ appearing in the denominator of Eq .(\ref{eq:tau_adv}) is performed in front of the shock surface and $v^r$ is the radial component of the three velocity $v^i$.
In the top right panel, we show spherical averaged shock (solid lines) and gain radii (dashed).
While in the rest of panels, to illustrate how the values vary relative to the rotational axis, we divide the space into two equal volume regions, polar and equator, and show the values evaluated in each region.
Here, we define the polar region (labeled by ``Pol'') by the cone angle of $60^\circ$ around the rotational axis, i.e., $\theta\le60^\circ$ or $\theta\ge120^\circ$, and the equatorial region  (labeled by ``Equ'') by $60^\circ<\theta<120^\circ$.
These ranges are used in the volume and surface integrals in Eqs. (\ref{eq:Mgain})-(\ref{eq:tau_heat}).
When we evaluate $\eta(=\dot Q/(L_{\nu_e}+L_{\bar \nu_e}))$, $\zeta$, and $\tau_{\rm adv}/\tau_{\rm heat}$, we first evaluate every quantity, e.g. $\dot Q$ and $L_{\nu_e}+L_{\bar \nu_e}$, in each region and then take their ratio.
Regarding the model R0B00, we show its values integrated over all solid angles (labeled by ``$4\pi$'') since it has basically no significant angle dependence.
Note that we show half values for extensive variables, i.e., $M_{\rm gain}$ and $\dot Q$, for model R0B00 to compare with other models.

Fig.~\ref{fig:TadvTheat} clearly shows how the rotational and magnetic field effects appear in general and also how they change the values relative to the rotational axis.
The (spherically averaged) gain radius locates more inward in rotating models R1B12 (red dashed line in the top right panel) and R1B00 (blue dashed) than the non-rotating model R0B00 (black dashed).
As can be seen in the top-left and -middle panels, the more inward $R_{\rm gain}$ and larger $R_{\rm shock}$ produce a more extended gain region and consequently a larger mass and total heating rate integrated over that region.
The non-rotating model R0B00 shows smallest $M_{\rm gain}$ and $\dot Q$, typically several times smaller than the other two.
The specific heating rate $\zeta$ (bottom left panel) also shows a rotational dependence.
In general, R0B00 presents higher $\zeta(=\dot Q/M_{\rm gain})$, although $M_{\rm gain}$ and $\dot Q$ themselves are smaller than the other two.
On the other hand, from the perspective of neutrino heating efficiency, $\eta(=\dot Q/(L_{\nu_e}+L_{\bar \nu_e}))$ in R0B00 shows the least efficiency (bottom middle).
Therefore rotation works to lower the specific heating rate $\zeta$ but raise the heating efficiency $\eta$.
Such a trend is consistent with previous rotating models with detailed neutrino transport in \cite{Summa18}.

In the bottom right panel, all these features mentioned above are aggregated in a value $\tau_{\rm adv}/\tau_{\rm heat}$.
Higher $\tau_{\rm adv}/\tau_{\rm heat}$ represents that the dwell time of matter in the gain region is relatively long in terms of heating time scale.
It thus leads to a more favorable condition for the explosion.
Particularly $\tau_{\rm adv}/\tau_{\rm heat}$ larger than one can be a measurement of the onset of runaway shock expansion due to neutrino-heating (see, \cite{BMuller17,Summa18,Ott18} for the latest 3D successful explosion models and also \cite{O'Connor18} for the 3D non-explosion models).
In the bottom right panel, model R1B12 which has the largest gain region shows highest $\tau_{\rm adv}/\tau_{\rm heat}$ (red lines), while model R0B00 shows the lowest value (black line).
Therefore our result also shows that rotation makes $\tau_{\rm adv}/\tau_{\rm heat}$ higher.
This tendency is again consistent with \cite{Summa18}.
In addition, the magnetic fields also assist the expansion of the shock surface and produce higher $\tau_{\rm adv}/\tau_{\rm heat}$ than the corresponding non-magnetized model R1B00.

Next we discuss how rotation affects the energetics in each region relative to the rotational axis.
First, in the model R1B00, both $M_{\rm gain}$ and $\dot Q$ show significantly higher values along the equator (blue dashed lines) than those in the polar region (blue solid).
The blue dashed and solid lines start to diverge when the second shock expansion takes place at $t_{\rm pb}\sim220$ ms.
The higher values seen in the equatorial region are again due to the rotational shock expansion.
These rotational effects were already discussed by \cite{Nakamura3D14}, though with a very simplified neutrino light bulb method, and we obtain a consistent result in our self-consistent M1 neutrino transport simulations.
The heating efficiency $\eta$ in the equatorial region is also nearly twice as high as that in the polar region.
As a consequence, $\tau_{\rm adv}/\tau_{\rm heat}$ exceeds one, only in the equator (blue dashed line), and not in the polar region (blue solid line).
If the neutrino heating were more efficient and could actually aid the second shock expansion, it would directly lead to the shock runaway phase.
The model R1B00, however, deflates and does not enter the runaway phase during our simulation time up to $t_{\rm pb}\sim500$ ms.

We see an interesting feature in model R1B12.
In this rotating magnetized model, as we have explained in Sec.~\ref{sec:Shock Wave Evolution}, it exhibits a rapid shock expansion toward the rotational axis soon after core bounce.
Therefore $\tau_{\rm adv}/\tau_{\rm heat}$ in the polar region (red solid line) shows slightly higher value than the equatorial one (red dashed).
However, the higher value in the red solid line only persists during the first $\sim100$ ms after bounce and afterward the red dashed line takes over the solid one with largely exceeding one.
Interestingly, $\tau_{\rm adv}/\tau_{\rm heat}$ in the polar region shows basically less than  unity till $t_{\rm pb}\sim200$ ms, although the shock runaway already occurs mainly toward the polar region.
The trend is thus completely opposite to that of R1B00 in which the region with larger shock expansion exhibits larger $\tau_{\rm adv}/\tau_{\rm heat}$.
We interpret these behaviors as that the neutrino heating is not the main mechanism of the bipolar shock expansion in R1B12, but the magnetic fields play the leading role to aid the shock expansion.
On the other hand, as the red dashed line is exceeding unity, the shock expansion along the equator is mainly supported by neutrino heating.

\subsection{The Asymmetry of Lepton Number Emission}
\label{sec:The Asymmetry of Lepton Number Emission}
\cite{Tamborra14ApJ} reported the existence of the lepton-number emission self-sustained asymmetry termed LESA.
This phenomenon is characterized by a spherical symmetry breaking of the lepton number emission, basically dominated by a dipole mode.
Their analysis exhibited that LESA appears together with a partial distribution of $Y_e$ in the PNS convection zone ($r\sim25$ km) suggesting that the partial distribution can possibly be the primary cause of LESA.
In their subsequent paper \citep{Glas19}, they also explained the origin of the partial distribution of $Y_e$ by the PNS convection.
They showed that the PNS convection excites preferentially the lower order multipole modes including the dipole one which drives partial distribution of $Y_e$.
In addition, once such a partial distribution of $Y_e$ is fully established, it results in a lepton number emission with a prominent dipole mode that heats more materials on the opposite side to the dipole mode enhancing a globally deformed shock surface.
Consequently, non-spherical mass accretion, basically with low mode $\ell=1$, on to the PNS core surface continues to replenish the lepton-rich matter and sustains the partial distribution of $Y_e$ \citep{Tamborra14ApJ}.
%%%%%%%%%%%%%%%%%%%%%%%%%%%%%%%%%%%%%%
\begin{figure}[htbp]
\begin{center}
\includegraphics[width=170mm,angle=-90.]{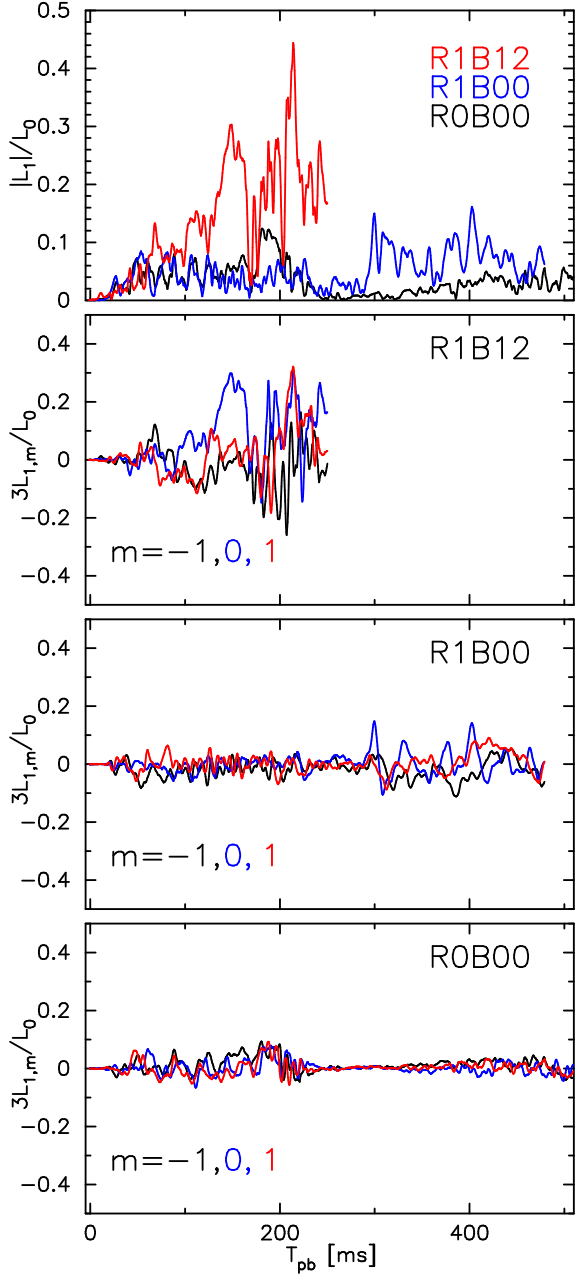}
  \caption{We plot the normalized dipole magnitude (top panel) and contributions from each quantum number $m$ for each model (bottom threes).
  In the top panel, the color represents the model name, while it indicates the quantum number $m=0,\pm1$ in other plots.
  We can see that the model R1B12 (red line in the top panel) shows a clear excess in its dipole magnitude.
  \label{fig:LESA}
}
\end{center}
\end{figure}
%%%%%%%%%%%%%%%%%%%%%%%%%%%%%%%%%%%%%%
%%%%%%%%%%%%%%%%%%%%%%%%%%%%%%%%%%%%%%

%Until recently, however, only one group \citep{Tamborra14ApJ} had reported the existence of LESA and the concern persisted that it could be a numerical artifact of the ray-by-ray-plus approach \citep{Buras06a}.
\cite{O'Connor18} and \cite{Vartanyan19} also reported the appearance of LESA using M1 neutrino transport method, i.e., full multi-dimensional neutrino transport.
\cite{O'Connor18} pointed out the importance of velocity dependent terms in the neutrino transport as the models without that term do not show any conclusive evidence for LESA.
\cite{Vartanyan19} also showed that the dipole mode can be comparable to the monopole one in the late post bounce phase $t_{\rm pb}\sim650$ ms.
Therefore, although the growth rate of dipole magnitude may actually depend on the detailed neutrino transport scheme \citep{Glas19}, the LESA seems to be a common phenomenon in CCSNe.

Following \cite{O'Connor18,Vartanyan19}, we plot the ratio of monopole to dipole mode of the lepton number emission as a function of the postbounce time in Fig.~\ref{fig:LESA}.
To plot the figure, we first evaluate the net lepton number flux via neutrinos $\mathcal{L}_{\nu}\equiv\mathcal{L}_{\nu_e}-\mathcal{L}_{\bar\nu_e}$ at $r=400$ km and then obtain the coefficient $\mathcal{L}_{\ell m}$ of spherical polar expansion of $\mathcal{L}_{\nu}$ as the same as what we do in Fig.~\ref{fig:Alm}.
In top panel, we plot the dipole magnitude $|\mathcal{L}_1|$ normalized by the monopole one $\mathcal{L}_0$, where we take the following definition \citep{O'Connor18}
\begin{eqnarray}
|\mathcal{L}_1|\equiv 3\sqrt{\sum_{m=-1}^1\mathcal{L}_{1m}^2}.
\end{eqnarray}
While in the lower three panels, we plot the value $3\mathcal{L}_{1m}/\mathcal{L}_0$ for each quantum number $m$ in each model to discuss the correlation with the shock morphology.

From top panel, we see that the absolute magnitude of normalized dipole mode in model R1B12 shows significantly larger value than the other two non-explosion models.
In this model R1B12, the dominant contribution to the total dipole mode is mainly coming from $m=0$ mode (blue line in the second panel).
Since it basically exhibits the positive value for $t_{\rm pb}\gtrsim100$ ms, the relative $\bar\nu_e$'s number flux is less toward positive $z$-axis and higher toward negative $z$-axis.
From Fig.~\ref{fig:Alm}, the shock morphology with $(\ell,m)=(1,0)$ mode becomes stronger for $t_{\rm pb}\gtrsim100$ ms with negative value that reflects that the shock expansion takes place relatively stronger toward the negative $z$-axis (also see, the final snapshot of the shock morphology in Fig.~\ref{fig:R1B12_Ent_Beta}).
It is thus opposite to the dipole mode of the lepton number flux.

\begin{figure*}[htbp]
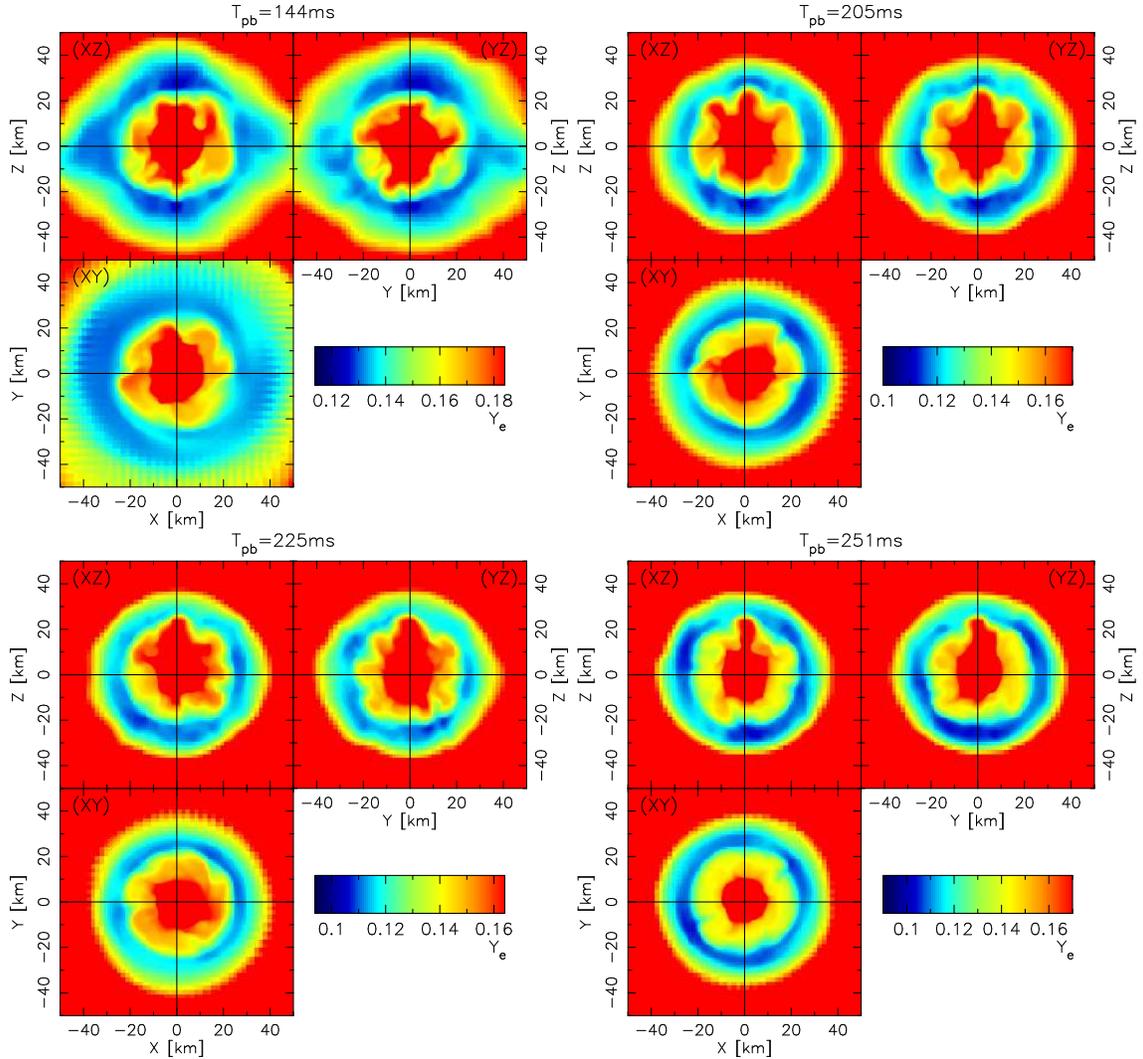

\begin{center}
\includegraphics[width=70mm,angle=-90.]{Ye1011.eps}
\includegraphics[width=70mm,angle=-90.]{Ye1311.eps}
\includegraphics[width=70mm,angle=-90.]{Ye1411.eps}
\includegraphics[width=70mm,angle=-90.]{Ye1544.eps}
  \caption{We show the distribution of $Y_e$ in the PNS at four different time slices $T_{\rm pb}=144$ (top-left), 205 (top-right), 225 (bottom-left), and 251 ms (bottom-right). In each panel, there are three minipanels that depict $xy$ (bottom-left), $xz$ (top-left), and $yz$ (top-right) planes. Although the partial distribution of $Y_e$ seemingly with higher order modes is visible at initial (say, $T_{\rm pb}=144$ and 205 ms, top two panels), we cannot see any clear dipole like structure. While, at around the end of our simulation time ($T_{\rm pb}\gtrsim220$ ms), the dipole mode seems to gradually grow with orienting toward positive $z$-axis. The structure is not destroyed by the PNS convection and persists at least for a few 10 milliseconds till the end of our simulation time in model R1B12.
  \label{fig:LESA_Ye_Distribution}
}
\end{center}
\end{figure*}

Although the anti-correlation between the orientation of the excess of the lepton number emission (positive $z$-axis) and the shock expansion (negative $z$-axis) seen in model R1B12 is consistent with the mechanism of LESA, the highest value $\sim0.4$ (red line in top panel) is significantly smaller than the values of \cite{Tamborra14ApJ}, in which they find the excess of dipole mode in all models irrespective of the explosion.
Therefore, to see if LESA is actually the mechanism of the excited dipole mode of the lepton number emission in model R1B12, we show in Fig.~\ref{fig:LESA_Ye_Distribution} the distribution of $Y_e$ in the PNS at four different time slices $T_{\rm pb}=144$ (top-left), 205 (top-right), 225 (bottom-left), and 251 ms (bottom-right).
In addition, we also show the distribution on $xy$ (bottom-left), $xz$ (top-left), and $yz$ (top-right) planes in every mini panel.
%%%%%%%%%%%%%%%%%%%%%%%%%%%%%%%%%%%%%%
%%%%%%%%%%%%%%%%%%%%%%%%%%%%%%%%%%%%%%
We note that, from the first and second panels in Fig.~\ref{fig:LESA}, the strong excess of lepton number emission mainly orienting toward positive $z$-axis is observed for $T_{\rm pb}\gtrsim100$ ms.
Therefore, if LESA is the origin of the excess, we would expect that the partial distribution of $Y_e$ has a dipole mode which orients opposite to the excess (see the schematic figure 15 in \cite{Tamborra14ApJ}), i.e., toward the positive $z$-axis in model R1B12.

From the $Y_e$ distribution at $T_{\rm pb}=144$ and 205 ms, we do not see any clear dipole like structure of $Y_e$ on $xz$ and $yz$ planes.
A clear dipole like structure appears only near the end of simulation time (see bottom two panels $T_{\rm pb}\gtrsim225$ ms).
%%%%%%%%%%%%%%%%%%%%%%%%%%%%%%%%%%%%%%
\begin{figure}[htbp]
\begin{center}
\includegraphics[width=50mm,angle=-90.]{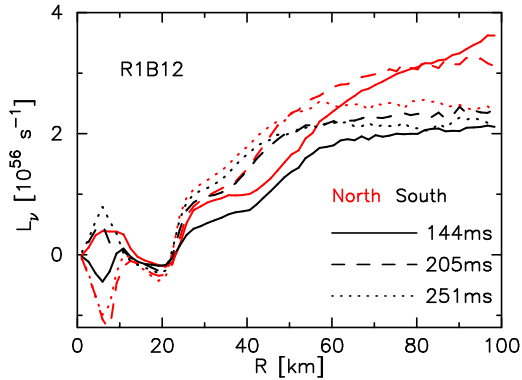}
  \caption{We plot the radial profile of the total lepton number flux $\mathcal{L}_{\nu}(=\mathcal{L}_{\nu_e}-\mathcal{L}_{\bar\nu_e})$ at three representative time slices $T_{\rm pb}=144$ (solid lines), 205 (dashed), and 251 ms (dotted). We plot $\mathcal{L}_{\nu}$ integrated over the northern (red lines) and southern (black) hemispheres at a given radius $R$.
  Here, from Fig.~\ref{fig:LESA}, the northern(southern) hemisphere corresponds to where we observe the excess(reduction) of $\mathcal{L}_{\nu}$, i.e., with relatively higher $\nu_e(\bar\nu_e)$ number flux.
  \label{fig:LESA_Lnu_Radial}
}
\end{center}
\end{figure}
%%%%%%%%%%%%%%%%%%%%%%%%%%%%%%%%%%%%%%
%%%%%%%%%%%%%%%%%%%%%%%%%%%%%%%%%%%%%%
The reason of the apparent inconsistency, namely the dipole lepton number emission without the existence of noticeable partial distribution of $Y_e$, can be understood from \cite{Tamborra14ApJ}.
According to their study, the dipole emission of total lepton number is produced mainly at two different regions, the PNS convection zone $R\lesssim20$ km, where the main dipole emission ($\sim70$-80 \%) occurs, and the entropy driven convection zone several 10 km $\lesssim R\lesssim R_{\rm shock}$ , where the dipole emission reaches its asymptotic value.
At the latter location, the partial distribution of $Y_e$ is established mainly by a replenishment of rich $Y_e$ material that is transported with stellar mantle deflected by the deformed shock surface.

Fig.~\ref{fig:LESA_Lnu_Radial} shows the radial profile of the total lepton number flux $\mathcal{L}_{\nu}(=\mathcal{L}_{\nu_e}-\mathcal{L}_{\bar\nu_e})$ at three representative time slices $T_{\rm pb}=144$ (solid lines), 205 (dashed), and 251 ms (dotted).
Here $\mathcal{L}_{\nu}$ is a hemispheric integration of the total lepton number flux measured in the comoving frame.
The hemispheric integration is performed for northern and southern hemispheres where we observe the excess and reduction of asymptotic $\mathcal{L}_{\nu}$, respectively.
By comparing the red and black lines at each time slice, they are almost overlapping just above the PNS convection zone $R\sim20$ km, while the difference gradually appears with radius especially at $R\gtrsim50$ km, i.e., in the entropy driven convection zone.
Such profile, namely the minor contribution from the PNS convection zone to the dipole emission, is completely different from the one in \cite{Tamborra14ApJ} which shows that the larger difference than ours already appears at $R\sim20$ km.
We, therefore, conclude that the dipole emission seen in model R1B12 is not originated from LESA but from the accretion induced partial distribution of $Y_e$ above the PNS convection zone.

%our value is comparable to those of \cite{O'Connor18} and \cite{Vartanyan19}.
%They reported the relative magnitudes of the dipole mode of $\sim0.5$ for $t_{\rm pb}\lesssim400$ ms.

 %As first pointed out by \citet{Tamborra14ApJ}, the LESA is driven by a replenishment of rich $Y_e$ material onto one side of the PNS core that can be established once the one-sided explosion morphology (note in our case, which is magnetorotationally-driven) becomes almost steady. Our results first show that the LESA originally proposed in the non-rotating progenitor also develops in our 3D MHD exploding model.

\section{Discussion and Conclusions}
\label{sec:Discussion and Conclusions}
We have presented the first 3D-GR MHD simulations of a 20 $M_\odot$ star with spectral neutrino transport. 
For the nuclear EOS and neutrino opacities, we used SFHo of \cite{SFH} and a baseline set of weak interactions \citep{Bruenn85,Rampp02}, where nucleon-nucleon bremsstrahlung is additionally taken into account, respectively.
Neutrino transport is handled by M1 closure scheme with the red and Doppler shift terms being fully considered.

We calculated three models, non-rotating non-magnetized, rotating non-magnetized, and rotating magnetized models to explore the effects of progenitor's rotation and magnetic field both on the dynamics and neutrino profiles.
Regarding the dynamics, while no shock revival was observed in two non-magnetized models during our simulation times, the shock expansion initiated shortly after bounce in a rotating magnetized model.
Initially the shock morphology takes a bipolar structure, which was eventually taken over by a unipolar one.
The shock front reached $1000$ km at $t_{\rm pb}\sim220$ ms and still continued expansion at the end of our simulation time.
From our analysis for the rotating magnetized model, we interpreted that the polar expansion is driven mainly by the magnetic pressure, while the equatorial expansion is facilitated by the neutrino heating.
Although we did not see the shock revival in two non-magnetized models, the standing shock locates further outward in the rotating model, which expands the gain region and increases the mass in the region.
Therefore, we obtained a consistent result with previous studies that the (moderate) rotation makes the condition more favorable for the explosion than the non-rotating case.

Using the same (or very similar) non-rotating 20 $M_\odot$ progenitor star as in this study, some previous 3D studies have shown a successful explosion \citep{Melson15b,Ott18,Burrows19}, while the others have not \citep{Tamborra14ApJ,Melson15b,O'Connor18}.
It is thus worth comparing our non-rotating and non-exploding model R0B00 with these previous studies.
One of major limitations in this work is its relatively lower numerical resolution compared to the previous ones.
It has been thoroughly examined that insufficient resolution can potentially inhibit the shock revival due to less turbulent pressure \citep[e.g.,][]{Couch15,BMuller15,Roberts16,Takiwaki16,Burrows19,Nagakura19}.
For instance, \cite{Ott18} performed full relativistic 3D calculations with M1 neutrino transport and obtained the shock revival.
This might be possibly due to their higher numerical resolution within the shock surface that achieves a factor of $\sim2-4$ higher than ours.
The higher numerical resolution allows the growth of turbulence leading to an additional pressure support.
It should be also noted that more up-to-date neutrino opacities, e.g., a strangeness-dependent contribution to the axial-vector coupling constant or many-body corrections to neutrino-nucleon scattering \citep{Burrows98,horowitz17}, generally benefit to facilitate the shock revival \citep[e.g.,][]{Kotake18,Burrows19}.
We are currently conducting CCSN simulations with better neutrino opacities following \cite{Kotake18}, which would be reported elsewhere in the near future.

We investigated the effect of the precollapse rotation and magnetic fields on the 
neutrino signals.
In general, both of the rotation and magnetic field decrease the neutrino luminosity and energy as they make the PNS core less compact due to the centrifugal force and/or mass ejection.
In addition, the rotation produces angle dependent neutrino signals relative to the rotational axis.
The neutrino luminosity and energy along the equator are significantly lower than those along the rotational axis.
%It stems from the rotational flattening of neutrino spheres, which moves the last scattering surface of neutrinos along the equatorial plane outward to the low temperature region.
We observed a quasi-periodic time modulation in the neutrino signals especially in model R1B00 toward the equator that is greatly suppressed along the rotational axis.
From our spectral analysis, the peak frequencies of the time modulated signals and of the $m=1$ deformation of neutrino sphere(s) have nearly the same value.
Therefore, together with the less modulation in heavier type neutrino signals, we consider that the spiral SASI mode deforms the neutrino spheres leading to the quasi-periodic signals.
Our results showed clear dependencies of neutrino signals on progenitor's rotation, magnetic field, and the observation angle. A more systematic study (such as changing the progenitor model, the initial magnetorotational strength, and the inclination between the rotation and magnetic axis) is needed for clarifying the multi-messenger signals from magnetorotationally-driven CCSNe.

We also witnessed the dipole emission of lepton number for our MR-explosion model, albeit weak. Although it is similar to the LESA phenomenon \citep{Tamborra14ApJ,O'Connor18,Glas19,Vartanyan19},
from our detailed analysis on the $Y_e$ distribution in the PNS convection layer and also on the spatial origin of dipole emission, we found that it is not associated with LESA.
We consider that the strong unipolar explosion in model R1B12 supplies rich $Y_e$ material on one side and produces the partial distribution leading to the dipole emission from the entropy driven convection zone. 
We, however, stress that more MHD simulations with sophisticated neutrino transport are indeed necessary to mention the robustness of the unipolar explosion seen in our model R1B12 and of the dipole emission associated with it.

As an important 3D effect, we showed that the kink instability is most likely to appear in the magnetized model that can potentially broaden the expanding blob, leading to weaker bipolar jets. %than those of the counterpart 2D model.
However, the PNS core may also be subject to the low-$T/W$ instability, we could not disentangle the outcomes of these two possibly coexisting instabilities.
Further numerical simulations by other independent groups, preferably with finer numerical resolutions, are definitely required to clarify the interplay between the two instabilities.

In the end of our discussion, we briefly mention the possible role of MRI.
Although the stellar magnetic field configuration and its strength at pre-collapse phase are poorly understood, strong initial magnetic fields $\sim10^{12}$ G as employed in this study might be too strong according to magnetized stellar evolution calculations by \cite{Heger05}, which gives $\lesssim10^{9}$ G \citep[but also see][for a possible scenario for considerably strong initial magnetic fields]{Soker19}.
To see how the MRI amplifies such plausibly weak magnetic fields, \cite{Obergaulinger09} conducted local shearing disk simulations.
Their results showed that the initial seed magnetic fields inside the PNS $\mathcal O(10^{12})$ G can be amplified to dynamically relevant strengths $\mathcal O(10^{15})$ G within several ms.
Since the main magnetic field amplification mechanism during core-collapse is compression, their initial seed magnetic fields inside the PNS $\mathcal O(10^{12})$ G could originate 
from the pre-collapse phase $\mathcal O(10^{9})$ G, which seems compatible with the stellar evolution calculation.

\cite{Sawai14} has shown in their global 2D axisymmetric simulations that the MRI can not only amplify the initial seed magnetic fields but also produce a global magnetic field in the postshock region. Later, a globally ordered field amplification in the PNS was found in full 3D-GR MHD simulations by \cite{Moesta14}.
Furthermore, \cite{Raynaud20} just recently reported the first numerical evidence of generation of magnetic fields inside the PNS convection zone with dynamically relevant strengths $\mathcal O(10^{15})$ G irrespective of the initial seed magnetic field strengths.
All these facts indicate that model R1B12 in this study is not too extreme but might be plausible, although the typical length scale of the MRI $\lesssim\mathcal{O}(10)$ m is far too small to resolve by our current numerical grid size (simply limited by our currently available computational resources).
Other than the MR explosion scenario, the turbulence in the MRI could enhance the neutrino heating efficiency, which could impact the neutrino mechanism \citep{Sawai14,Masada15}. All these subjects remain to be studied. As such, we can see a vast untouched (research) territory lying in front of us, into which we have just made a first jump with a newly developed tool (our 3D-GR MHD code) in hand.

\acknowledgements{
We thank Shota Shibagaki, Martin Obergaulinger, and Federico Maria Guercilena for helpful discussions and useful comments.
We also acknowledge H.-T. Janka for his valuable comments on the dipole emission of lepton number.
This research was supported by the ERC Starting Grant EUROPIUM-677912 (TK and AA),
JSPS KAKENHI Grant Number (JP17H05206, JP17K14306, and
JP17H01130, JP17H06364, JP18H01212 (KK and TT)),
and JICFuS as a priority issue to be tackled by using the Post `K' Computer.
Numerical computations were carried out on Cray XC50 at CfCA, NAOJ.
}

\begin{figure*}[htbp]
\begin{center}
\includegraphics[width=60mm,angle=-90.]{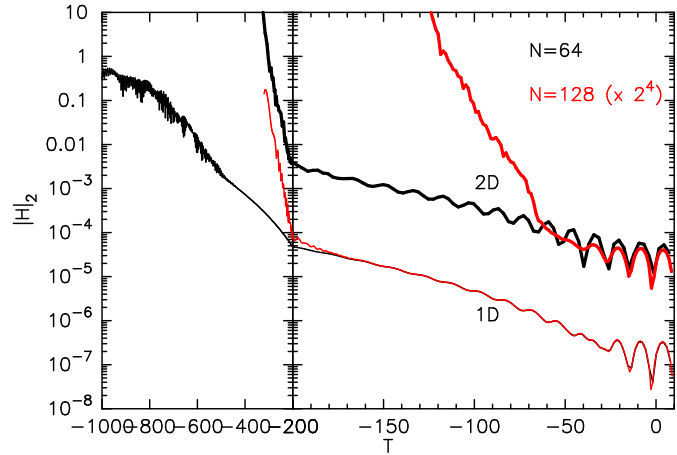}
  \caption{We plot the $L_2$ norm of violation of the Hamiltonian constraint $|\mathcal{H}|_2$ for coarser spacing model with $N=64$ (black lines) and finer ones with $128$ (red).
  For finer resolution models (red lines), we multiply $|\mathcal{H}|_2$ by $2^4$, since we use fourth-order spatial finite differencing.
  We evolve the metric backward in time starting at $T\sim9.875$.
  \label{fig:Gowdy}
}
\end{center}
\end{figure*}

\appendix
In this appendix, we show that our metric evolution implementation has a fourth-order convergence in space by checking the well-known polarized Gowdy wave test \citep{Alcubierre04}.
We omit to write the Gowdy wave metric and initial condition that can be found elsewhere \citep[e.g.,][]{Alcubierre04}.
We evolve the collapsing Gowdy-wave metric backwards in time using the harmonic slicing condition with zero shift vector $\beta^i=0$ as for the gauge condition.
Although the Gowdy-wave is a plane wave, we perform the test both in full 1D and 2D space.
In the latter 2D case, we tilt the propagation direction of the plane wave at 45$^\circ$ in the $xy$-plane.
We employ two different grid spacing $dx=1/N$ with $N=64$ or $128$ to check for the numerical convergence.
Fig.~\ref{fig:Gowdy} shows the $L_2$ norm of violation of the Hamiltonian constraint $|\mathcal{H}|_2$ for coarser spacing model with $N=64$ (black line) and finer one with $128$ (red).
For finer resolution models (red lines), we multiply $|\mathcal{H}|_2$ by $2^4$, since we use fourth-order spatial finite differencing.
From the figure, we see that there is almost a perfect overlap during the first $\sim180$ and $\sim40$ crossing times in 1D and 2D test, respectively, which shows that our metric evolution scheme actually achieves a fourth-order convergence in space.
\label{sec:Gowdy}
%%%%%%%%%%%%%%%%%%%%%%%%%%%%%%%%%%%%%%
%%%%%%%%%%%%%%%%%%%%%%%%%%%%%%%%%%%%%%

\bibliographystyle{apj}
\bibliography{mybib}

\end{document}